\documentclass[magazine]{IEEEtran}
\usepackage{setspace}

\usepackage{url,mathdots,algorithmic,bm}            
\usepackage{booktabs,comment}       
\usepackage{amsfonts}       
\usepackage{microtype}      
\usepackage{amsmath,amsbsy,amstext,amssymb}
\usepackage{color}
\usepackage{algorithm,times}
\usepackage{subfigure}
\usepackage{latexsym,graphicx,psfrag,wrapfig,breqn}
\usepackage{booktabs,xspace}

\graphicspath{{figures/}}

\usepackage{float}
\floatstyle{boxed}
\newfloat{textbox}{thp}{lop}
\usepackage{capt-of}

\ifCLASSINFOpdf
\else
\fi
\title{Harnessing Sparsity over the Continuum: \\
Atomic Norm Minimization for Super Resolution}

\author{Yuejie Chi,~\textit{Senior Member, IEEE}, Maxime Ferreira Da Costa,~\textit{Member, IEEE}
\thanks{The authors are with the Department of Electrical and Computer Engineering, Carnegie Mellon University, Pittsburgh, PA, USA (emails: \{yuejiechi, mferreira\}@cmu.edu).}}

\newcommand{\cA}{{\mathcal{A}}}

\newcommand{\bc}{{\boldsymbol c}}

\newcommand{\bz}{{\boldsymbol z}}

\newcommand{\bu}{{\boldsymbol u}}

\newcommand{\bA}{{\boldsymbol A}}
\newcommand{\bG}{{\boldsymbol G}}

\newcommand{\ba}{{\boldsymbol a}}
\newcommand{\bb}{{\boldsymbol b}}
\newcommand{\bw}{{\boldsymbol w}}

\newcommand{\bx}{{\boldsymbol x}}

\newcommand{\by}{{\boldsymbol y}}
\newcommand{\br}{{\boldsymbol r}}

\newcommand{\bg}{{\boldsymbol g}}

\newcommand{\bX}{{\boldsymbol X}}
\newcommand{\bH}{{\boldsymbol H}}

\newcommand{\bW}{{\boldsymbol W}}

\newcommand{\bs}{{\boldsymbol s}}

\newcommand{\bP}{{\boldsymbol P}}

\newcommand{\bp}{{\boldsymbol p}}

\newcommand{\argmax}{\mathop{\rm argmax}}
\newcommand{\argmin}{\mathop{\rm argmin}}

\newcommand{\sgn}{\mathop{\rm sgn}}

\newcommand{\conv}{\mathop{\rm conv}}

\newcommand{\toep}{\mathop{\rm toep}}

\newcommand{\tr}{\mathop{\rm Tr}}

\newcommand{\bbC}{\mathbb{C}}

\newcommand\anorm[1]{\left\Vert #1 \right\Vert_{\cA}}

\renewcommand{\Re}{\operatorname{Re}}

\begin{document}

\maketitle


\IEEEdisplaynontitleabstractindextext

At the core of many sensing and imaging applications, the signal of interest can be modeled as a linear superposition of translated or modulated versions of some template (e.g. a point spread function, a Green's function), and the fundamental problem is to estimate the translation or modulation parameters (e.g. delays, locations, Dopplers) from noisy measurements. This problem is of central importance to not only target localization in radar and sonar, channel estimation in wireless communications, direction-of-arrival estimation in array signal processing, but also modern imaging modalities such as super-resolution single-molecule fluorescence microscopy, nuclear magnetic resonance imaging, spike localization in neural recordings, among others.

Typically, the temporal or spatial resolution of the acquired signal is limited by that of the sensing or imaging devices, due to factors such as the numerical aperture of a microscope, the wavelength of the impinging electromagnetic or optical waves, or the sampling rate of an analog-to-digital converter. This resolution limit is well-known and often referred to as the {\em Rayleigh limit} (c.f. the box ``What is the Rayleigh Limit?''). The performance of matched filtering, or periodogram, which creates a correlation map of the acquired signal against the range of parameters, is limited by the Rayleigh limit, regardless of the noise level.

On the other hand, the desired resolution of parameter estimation can be much higher -- a challenge known as {\em super resolution}. There is a long history of pursuing super-resolution algorithms in the community of signal processing \cite{stoica1997introduction,kay1981spectrum,krim1996two}. The oldest one probably dates back to de Prony's root finding method in as early as 1795 \cite{prony1795essai}, and variants of this method that are better suited for noisy data have also been proposed over time, see e.g.   \cite{kumaresan1984prony}. Subspace methods based on the computation of eigenvector or singular vector decompositions, such as MUSIC \cite{Schmidt1986MUSIC}, ESPRIT \cite{RoyKailathESPIRIT1989} and matrix pencil \cite{HuaSarkar1990}, are another class of popular approaches since their inception in 1980s. Different forms of maximum likelihood estimators have also been studied extensively \cite{stoica1989maximum,clark1994two}. Collectively, these algorithms have ``super resolution'' capabilities, namely, they can resolve the parameters of interest at a resolution below the Rayleigh limit when the noise level is sufficiently small.

\begin{textbox}[ht]
\begin{center}
{\bf What is the Rayleigh Limit?} \\
\end{center}

The Rayleigh limit is an {\em empirical} criterion characterizing the resolution of an optical system due to diffraction. Taking a conventional fluorescence microscope as an example, the observed diffraction patterns of two fluorescent point sources become visually harder to distinguish as the point sources get closer to each other, illustrated below. We say they are no longer resolvable when their separation is below the Rayleigh limit ($\mathsf{RL}$).
\begin{center}
    \label{fig:rayleighLimit}
    \centering
    \includegraphics[width=0.9\columnwidth]{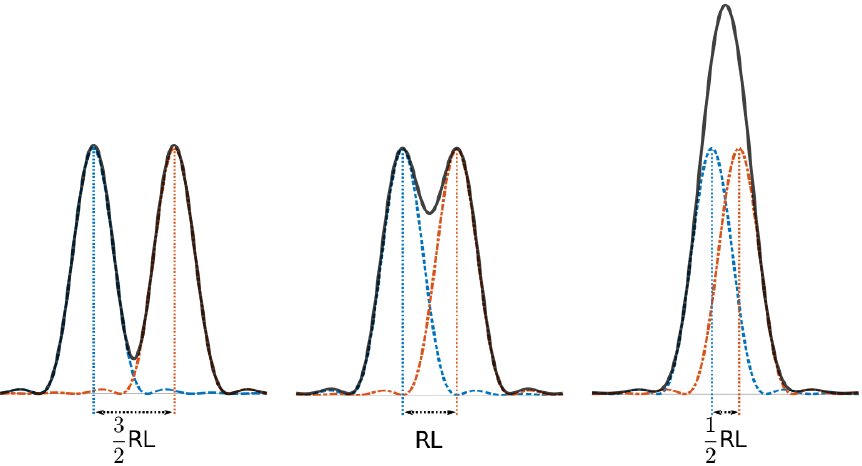}
    \captionof{figure}{The combined response for two translated point spread functions under different separations of the point sources. The $\mathsf{RL}$ is an indication of the separability of the two sources.}
\end{center}

\end{textbox}

While there already exists a plethora of traditional methods,  convex optimization recently emerges as a compelling framework for performing super resolution, garnering significant attention from multiple communities spanning signal processing, applied mathematics, and optimization. Due to (the relative) tractability of convex analysis and convex optimization, the new framework offers several benefits. First, strong theoretical guarantees are rigorously established to back up its performance even in the presence of noise and corruptions. Second, it is versatile to include prior knowledge into the convex program to handle a wide range of measurement models that fall out of the reach of traditional methods. Third, leveraging the rapid progress in large-scale convex optimization, it opens up the possibility of applying efficient solvers tailored to real-world applications.

The goal of this paper is to offer a friendly exposition to atomic norm minimization \cite{CandesFernandez2012SR} as a {\em canonical} convex approach for super resolution. The atomic norm is first proposed in \cite{chandrasekaran2012convex} as a general framework for designing tight convex relaxations to promote ``simple'' signal decompositions, where one seeks to use a minimal number of ``atoms'' to represent a given signal from an atomic set composed of an ensemble of signal atoms. Celebrated convex relaxations such as the $\ell_1$ norm approach for cardinality minimization \cite{CheDonSau01} and the nuclear norm approach for rank minimization \cite{recht2010guaranteed} can be viewed as particular instances of atomic norms for appropriately defined atomic sets. Specializing the atomic set to a dictionary containing all translations of the template signal over the continuous-valued parameter space, estimating the underlying translation parameters is then equivalent to identifying a sparse decomposition in an infinite-dimensional dictionary. This key observation allows one to recast super resolution as solving an infinite-dimensional convex program \cite{tang2012compressed} -- a special form of atomic norm minimization considered in this paper. We first highlight its mathematical formulation through a pedagogical yet useful model of super resolution that amounts to line spectrum estimation, where this infinite-dimensional convex program can be equivalently reformulated as a semidefinite program, then demonstrate its versatility by discussing how it can be adapted to address measurement models that traditional methods may not be easily applicable. Finally, we illustrate its utility in super resolution image reconstruction for single-molecule fluorescence microscopy \cite{huang2009super}, where the infinite-dimensional convex program can be solved efficiently via tailored solvers. 

Throughout this paper, we use boldface letters to represent matrices and vectors, e.g. $\ba$ and $\bA$. We use $\bA^\top$, $\bA^{\mathsf{H}}$, $\tr(\bA)$ to represent the transpose, Hermitian transpose, and trace of $\bA$, respectively. The conjugate of a complex scalar $a$ is denoted as $a^*$. We use $\bA\succeq 0$ to represent $\bA$ is positive semidefinite. The matrix $\toep\left(\bu\right)$ denotes the Hermitian Toeplitz matrix whose first column is equal to $\bu$, and $\mbox{diag}(\bm{g})$ denotes the diagonal matrix with diagonal entries given as $\bm{g}$. The inner product between two matrices $\bX$ and $\bP$ is defined as $\left\langle \bX, \bP \right\rangle = \tr(\bX^{\mathsf{H}}\bP)$. Additionally, the notation $f(n)=O\left(g(n)\right)$ means that there exists a constant $c>0$ such that $\left|f(n)\right|\leq c|g(n)|$.

\section{What is the Atomic Norm?}\label{sec:AtomicNorm-Introduction}
 
An everlasting idea in signal processing is decomposing a signal into a linear combination of judiciously chosen basis vectors, and seeking compact and interpretable signal representations that are useful for downstream processing. For example, decomposing time series into sinusoids, speeches and images into wavelets, total system responses into impulse responses, etc.
 
To fix ideas, consider the task of representing a signal $\bx$ in a vector space using {\em atoms} from a collection of vectors in  $\mathcal{A}=\{\ba_i \}$ called an {\em atomic set}. The set $\cA$ can contain either finite or infinite numbers of atoms. We wish to expand $\bx$ using the atoms in a form of
\begin{equation}\label{eq:abstract-atomic-decomposition}
    \bx = \sum_i c_i \ba_i, \quad \ba_i \in\cA,
\end{equation}
where $c_i > 0 $ specifies the coefficients of the decomposition. In many applications, the size of $\cA$ can be much larger than the dimension of the signal, leading to an overcomplete representation, and there are an infinite number of possibilities to decompose $\bx$. Which representation, then, shall we pick? Among the many plausible criteria, one meaningful approach is to pursue the Occam's razor principle, and seek for a parsimonious decomposition of the signal $\bx$ involving the \emph{smallest possible number of atoms} in $\cA$, i.e. the sparsest solution to \eqref{eq:abstract-atomic-decomposition}. The corresponding representation is known as a {\em sparse representation} \cite{DonohoHuo01}. Many real-world signals admit sparse representations for appropriately chosen atomic sets. 
As a simple example, natural images are approximately sparse by picking $\mathcal{A}$ as a wavelet frame. Low-rank matrices, another class of signals that have enjoyed wide success in signal processing \cite{chen2018harnessing}, are sparse with respect to an atomic set $\mathcal{A}$ that is the collection of all unit-norm rank-one matrices.

Given a signal $\bx$, how to find its sparse representation in the atomic set $\cA$? In general, this problem is nonconvex and can be NP-hard due to the combinatorial aspect of cardinality minimization. The key motivation behind atomic norm minimization, proposed by Chandrasekaran et al. \cite{chandrasekaran2012convex}, is to relax the nonconvex sparsity cost by its tight convex surrogate, and solve instead the resulting convex relaxation that is more tractable. This idea is a generalization of the popular $\ell_1$ minimization for sparse vector recovery \cite{CandesRombergTao2006Stable,Donoho06} when $\cA$ is a {\em finite} set. Therein, one seeks to solve a linear program which minimizes the sum instead of the cardinality of the nonzero coefficients.

\begin{figure}[t]
    \centering
    \includegraphics[width=0.8\columnwidth]{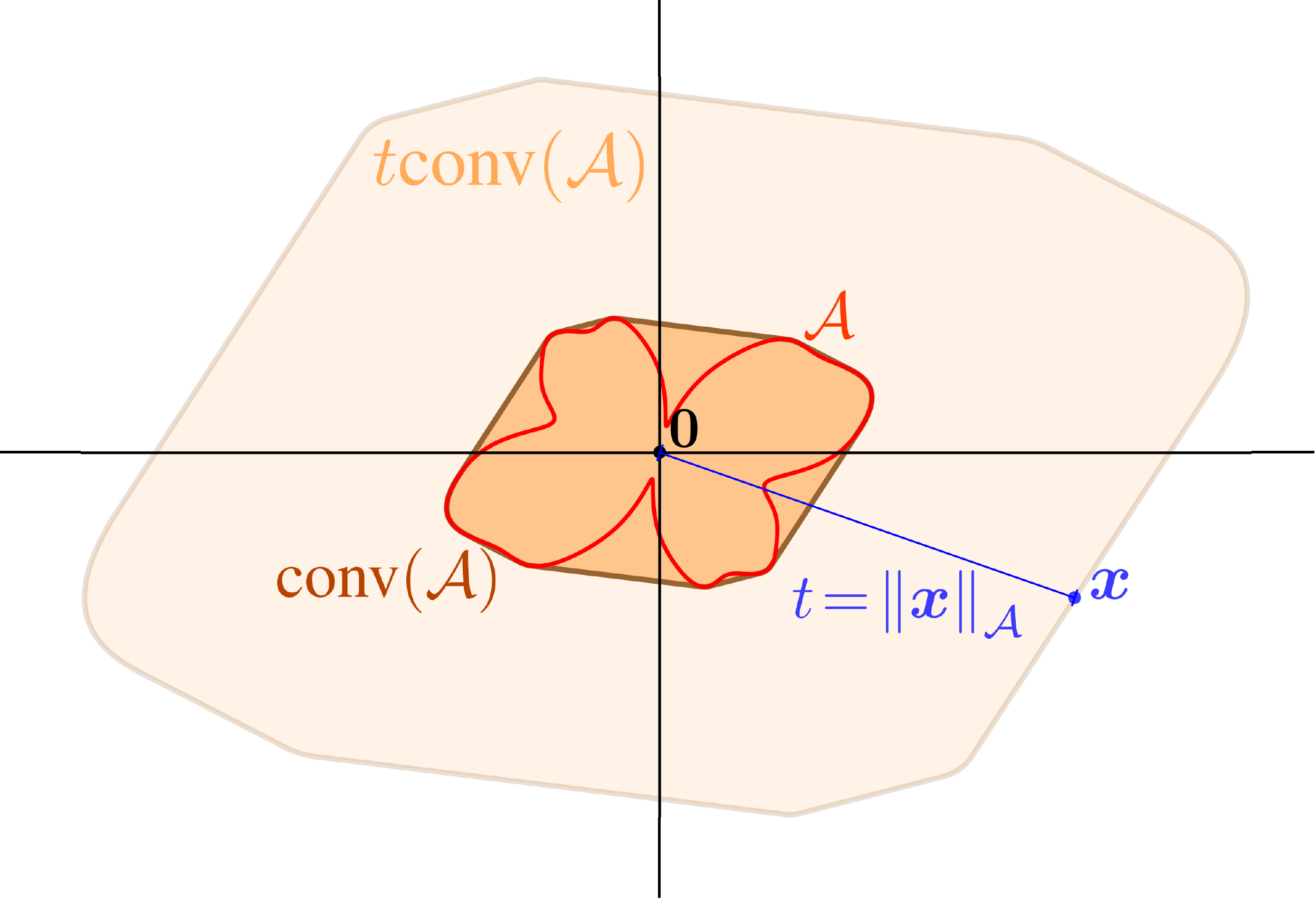}
    \caption{An atomic set $\cA$ (in red) and its convex hull $\conv (\cA)$ (in orange). The atomic norm of a vector $\bx$ can be interpreted as the smallest dilation factor $t\geq 0$ such that $\bx$ belongs to  $t  \conv (\cA)$ (in blue).}\label{fig:atomic_norm}
\end{figure}

To extend the same idea to the case where $\cA$ is an arbitrary, and possibly infinite-dimensional set, we first take the convex hull of $\cA$, denoted as $\conv(\cA)$, and then define its associated Minkowski functional (or gauge function) as \cite{chandrasekaran2012convex}
\begin{equation}\label{eq:atomic-norm-definition}
	\anorm{\bx} \triangleq \inf \left\{ t \geq 0: \; \bx \in t \cdot \conv \left(\cA \right) \right\},
\end{equation}
which is the solution to a convex program. When $\cA$ is centrally symmetric about the origin, the above definition leads to a valid norm, and is called the {\em atomic norm} of $\bx$. Fig.~\ref{fig:atomic_norm} presents an illustration of this concept, where the atomic norm is the smallest nonnegative scaling of $\conv(\cA)$ until it intersects $\bx$. Following the definition \eqref{eq:atomic-norm-definition}, a fundamental geometric property is that the atomic norm ball, i.e., $\{\bx: \; \|\bx\|_{\cA}\; \leq 1 \}$, is exactly $\conv(\cA)$. 

More interestingly, consider the case when $\bx$ lies in an $n$-dimensional vector space. Carath\'eodory's theorem \cite{caratheodory1911variabilitatsbereich} guarantees that any point in $\conv(\cA)$ can be decomposed as a convex combination of at most $n+1$ points in $\cA$, where $\cA$ is not necessarily convex. Therefore, one may rewrite \eqref{eq:atomic-norm-definition} as
\begin{equation}\label{eq:atomic-norm-decomposition}
	\anorm{\bx} = \inf \left\{\sum_i  c_i  : \; \bx=\sum_i c_i \ba_i,\;c_i > 0,\; \ba_i\in\cA \right\},
\end{equation}
as long as the centroid of $\conv(\cA)$ is the origin. The decomposition $\sum_i c_i \ba_i$ that obtains the infimum is referred to as the \emph{atomic decomposition} of $\bx$ onto $\cA$. It is not hard to see that the atomic norm indeed subsumes the $\ell_1$ norm as a special case but accommodates the more general case where $\cA$ can be an infinite-dimensional set.

Several central questions are how to properly select the atomic set, compute the atomic norm and find the atomic decomposition, and when the atomic decomposition coincides with the sparse representation, i.e. the convex relaxation is tight. Clearly, the answers depend highly on the atomic set as well as the signal itself. These questions have been addressed extensively in the study of $\ell_1$ norm minimization for sparse vector recovery  \cite{CandesRombergTao2006Stable,Donoho06,foucart2013mathematical}. In the context of super resolution, we will first address these questions under a simple model that amounts to the classical problem of line spectrum estimation, which has deep connections to systems and control theory.

\begin{figure*}[t]
    \centering
    \includegraphics[width=0.9\textwidth]{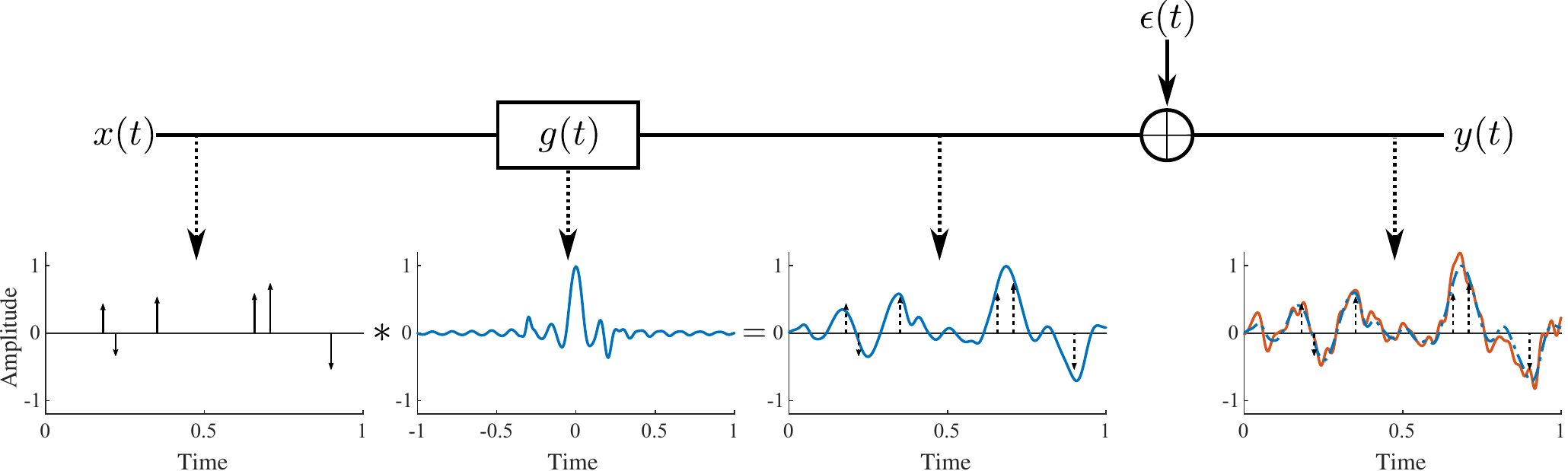}
    \caption{An illustration of the mathematical model of super resolution. The spike signal $x (t )$ is convolved with a point spread function $g (t )$, leading to degradation of its resolution, which is further exacerbated by an additive noise $\epsilon (t )$, producing an output signal $y (t )$.} \label{fig:observationModel}
\end{figure*}

\section{A Mathematical Model of Super Resolution, Equivalent to Line Spectrum Estimation}
\label{sec:model}
We first focus on a simple yet widely applicable model of super resolution  that describes the convolution of a sequence of point sources with a point spread function (PSF) that is resolution-limited, illustrated in Fig.~\ref{fig:observationModel}. Let $x(t)$ be a spike signal given as
\begin{equation}\label{eq:observation-model}
    x(t) = \sum_{k=1}^{r} c_{k}\delta(t-\tau_{k}).
\end{equation}
Here, $r$ is the number of spikes, $c_k\in\mathbb{C}$ and $\tau_k\in [0,1)$\footnote{Without loss of generality, the maximal delay is normalized to $1$.} are the complex amplitude and delay of the $k$th spike. Such a spike signal can model many physical phenomena, such as firing times of neurons, locations of fluorescence molecules, and so on. Let $g(t)$ be the PSF whose bandwidth is limited due to the Rayleigh limit, namely its Fourier transform $G(f)$ satisfies
$$ G(f) = 0 \quad \mbox{whenever}\quad |f| > B/2$$
for some bandwidth $B>0$. Its convolution with $x(t)$, contaminated by an additive noise $\epsilon(t)$, can be written as 
\begin{align*} 
y(t) &=   x( t )* g(t) + \epsilon(t)  =  \sum_{k=1}^{r} c_{k} g(t-\tau_{k}) + \epsilon(t),  
\end{align*}
where $*$ denotes the convolution operator. Sampling the Fourier transform of the above equation at the frequencies $\ell=- \lfloor B/2 \rfloor, \cdots, 0, \cdots , \lfloor B/2 \rfloor$, we obtain the measurements
\begin{equation}\label{eq:receive_freq}
    Y_\ell = G_\ell \cdot X_\ell + E_\ell = G_\ell \cdot \left( \sum_{k=1}^{r} c_{k} e^{-j2\pi \ell \tau_k} \right) + E_\ell,
\end{equation}
where $G_\ell$, $X_\ell$, $E_\ell$, and $Y_\ell$ are the Fourier transforms of $g(t)$, $x(t)$, $\epsilon(t)$, and $y(t)$ evaluated at frequency $\ell$, respectively. The total number of samples is $n = 2\lfloor B/2 \rfloor +1 \approx B$. We write \eqref{eq:receive_freq} in a vector form as
\begin{equation}\label{eq:samples_vector}
    \bm{y} = \mbox{diag}(\bm{g}) \bm{x} + \bm{\epsilon},
\end{equation}
where $\bm{y} =[Y_\ell]$, $\bm{g}=[G_\ell]$, $\bm{x}=[X_\ell]$, and $\bm{\epsilon}=[E_\ell]$. The problem of super resolution is then to estimate $\{c_k,\tau_k\}_{1\leq k\leq r}$ accurately from $\bm{y}$, without knowing the model order $r$ a priori. Here, the Rayleigh limit is inversely proportional to the bandwidth $B$, and roughly speaking, is about $1/n$.

When the PSF $g(t)$ is known, one can ``equalize'' \eqref{eq:receive_freq} by multiplying $G_\ell^{-1}$ to both sides, provided that $G_\ell$'s are non-zero. The observation $\bz =[G_\ell^{-1}Y_{\ell}]$ relates to $\bx$ as
\begin{equation}\label{eq:model_withoutPSF}
    \bm{z}  = \bm{x} + \widetilde{\bm{\epsilon}},
\end{equation}
where $ \widetilde{\bm{\epsilon}} $ is the additive noise. With slight abuse of notation, we map the index of $\ell$ from $\lfloor B/2 \rfloor,\cdots, -\lfloor B/2 \rfloor$ to $0,\cdots, n-1$ for convenience, and write $\bm{x}$ as a superposition of complex sinusoids: 
\begin{equation}\label{eq:single_mv}
\bx= \sum_{k=1}^r c_{k} \ba(\tau_k),
\end{equation}
where $\ba\left(\tau \right)\in\bbC^n$ is a vector defined as 
\begin{equation} \label{eq:spectral_atom}
\ba (\tau) =  \left[1, e^{j2\pi \tau}, \ldots, e^{j2\pi (n-1)\tau}\right]^\top, \quad \tau \in [0,1).
\end{equation}

Notably, the above simplified model \eqref{eq:model_withoutPSF} also amounts to the classical problem of {\em line spectrum estimation}, that consists of estimating a mixture of sinusoids (with frequencies $\tau_k\in[0,1)$) from equi-spaced time samples (sampled at integers $\{0,\cdots, n-1\}$) of the time-domain signal $x_{\mathsf{ls}}(t) = \sum_{k=1}^{r} c_k e^{j2\pi \tau_k t}$. This finds applications in speech processing, power system monitoring, systems identification, and so on.  The same model also describes direction-of-arrivals estimation using a uniform linear array, which is studied extensively in the literature of spectrum analysis \cite{stoica1997introduction}.

\section{Line Spectrum Super Resolution via Atomic Norm Minimization}

\begin{figure*}[t]
    \centering
    \begin{tabular}{cccc}
    \includegraphics[width=0.212\textwidth]{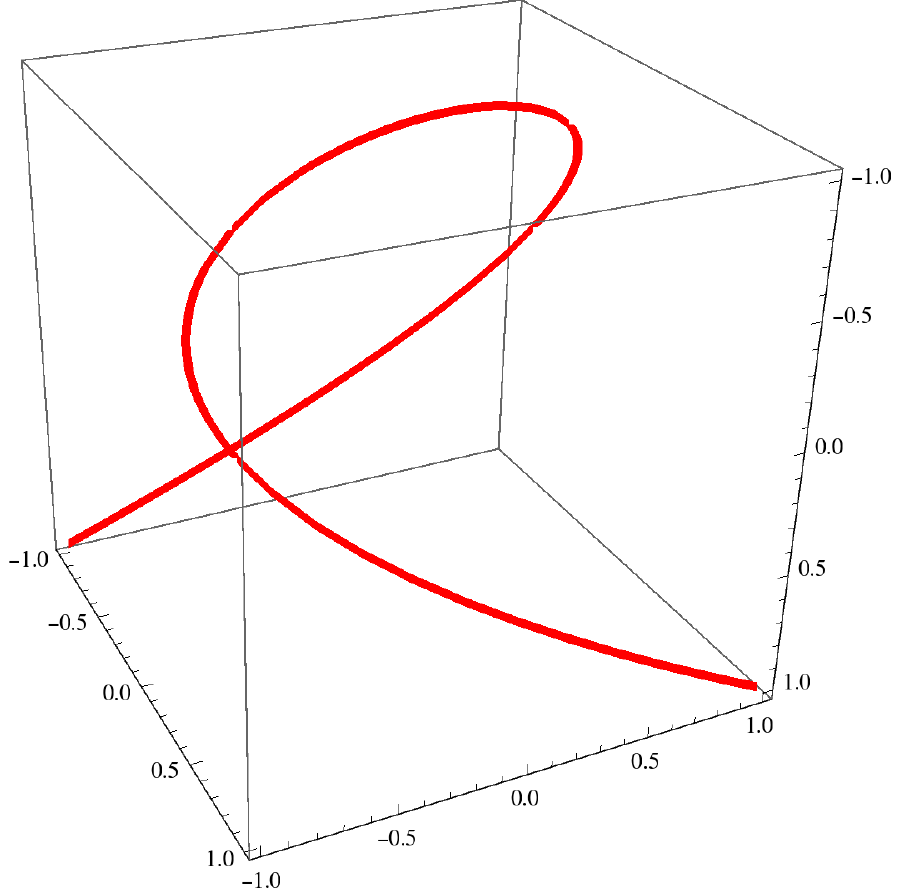}\hspace{0.1in} & \includegraphics[width=0.212\textwidth]{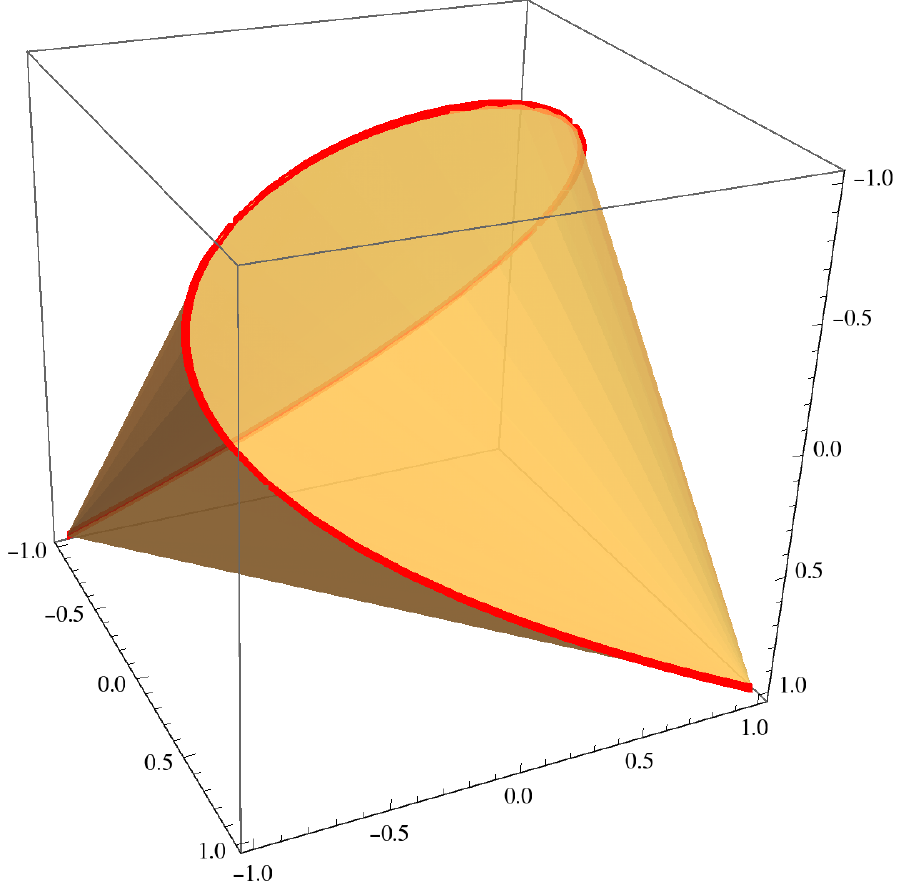} \hspace{0.1in} &
    \includegraphics[width=0.212\textwidth]{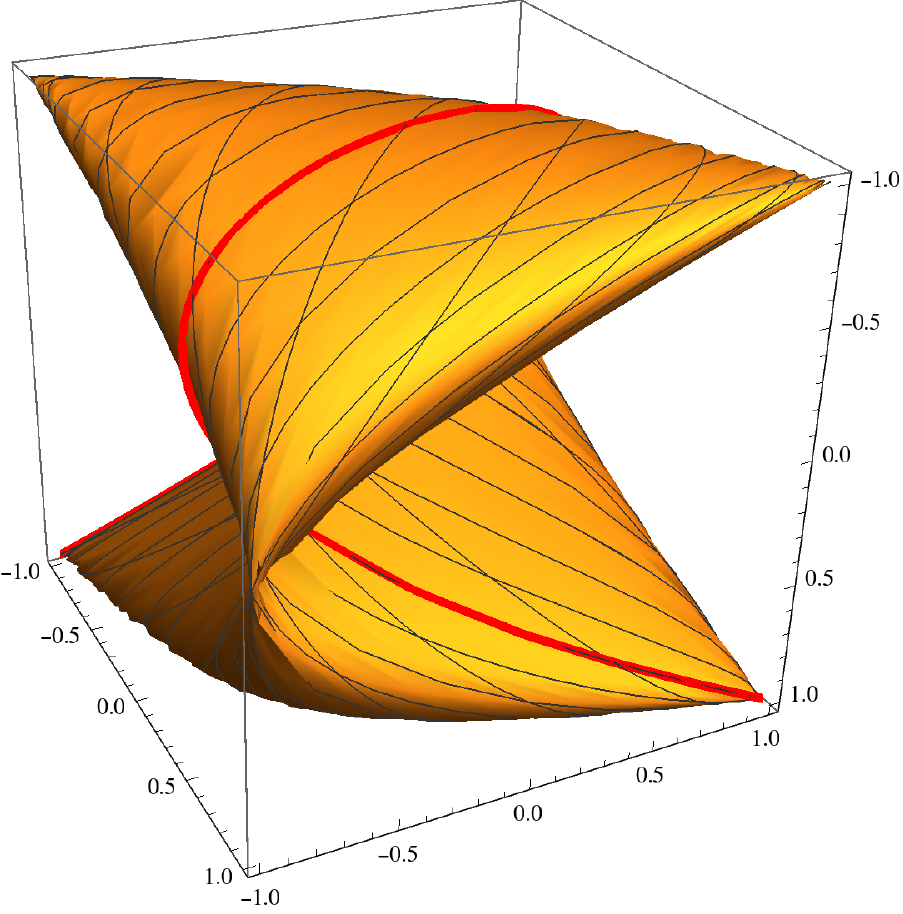}\hspace{0.1in}  &
    \includegraphics[width=0.212\textwidth]{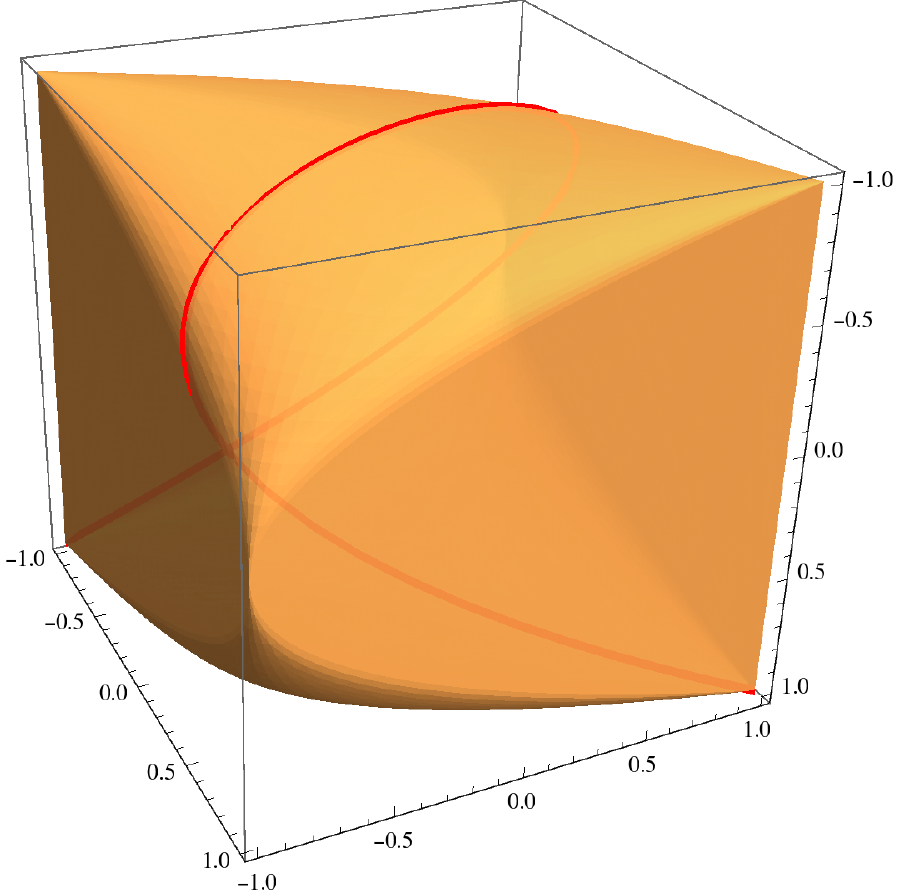}\\
    (a) & (b) & (c) & (d)
    \end{tabular}
    \caption{Visualization of the continuous-valued atomic set for line spectrum super resolution. (a) The moment curve $\cA_0$ restricted to the 3 real moments $\{\left[\cos(2\pi\tau),\cos(4\pi\tau),\cos(6\pi\tau)\right]^\top:\;\tau\in\left[0,1\right)\}$, and (b) its convex hull.  (c) The phased version of the moment curve $\cA_{\mathsf{1D}}$ restricted to the 3 real moments $\{\left[\cos(2\pi\tau+\phi),\cos(4\pi\tau+\phi),\cos(6\pi\tau+\phi)\right]^\top:\;\tau\in\left[0,1\right), \phi\in [0,2\pi)\}$, and (d) its convex hull.}
    \label{fig:momentCurve}
\end{figure*}

In the absence of noise, one could think of super resolution as estimating the continuous-time spike signal $x(t)$ in \eqref{eq:observation-model} from its discrete-time moment measurements $\bx$ in \eqref{eq:single_mv}, which are related through
\begin{equation}\label{eq:integral_form}
\bx = \int_0^1 \ba(t) \mathrm{d}x(t).
\end{equation}
One can also think of $x(t)$ as the representation of $\bm{x}$ over a continuous dictionary
\begin{equation}\label{eq:moment_curve}
    \cA_{0}= \left\{\ba(\tau):\;\tau\in\left[0,1\right)\right\},
\end{equation}
which forms a one-dimensional variety of $\mathbb{C}^{n}$ called the \emph{moment curve}, illustrated in Fig.~\ref{fig:momentCurve} (a). It is well-known that the convex hull of
$\cA_0$, illustrated in Fig.~\ref{fig:momentCurve} (b), is a body of $\mathbb{C}^n$ that can be parameterized by a set of linear
matrix inequalities \cite{sanyal2011orbitopes}, and has close relationships with the positivity of Hermitian Toeplitz matrices. This fundamental property of the moment curve has many implications in control and signal processing \cite{lasserre2010moments,dumitrescu2007positive}, and is key to the development of a super-resolution theory based on atomic norm minimization.

It is clearly possible to obtain the same $\bx$ from different $x(t)$. However, if we impose some sparsity assumption, namely constraining how many spikes are allowed in $x(t)$, this representation can be ensured to be unique. In particular, the representation \eqref{eq:single_mv} is unique as long as $r\leq \left\lfloor n/2 \right \rfloor$ and the support set $\mathcal{T}=\{\tau_k\}_{1\leq k\leq r}$ contains distinct elements.

\subsection{Atomic Norm for Line Spectrum Super Resolution}

To apply the framework of atomic norm minimization for super resolution, one must first define the atomic set properly. Since the complex amplitudes $c_k$'s can take arbitrary phases, we introduce an augmented atomic set taking this into account:
\begin{equation}\label{eq:atomic_set_sr}
\cA_{\mathsf{1D}}=\left\{ e^{j\phi} \ba( \tau ) : \; \tau\in \left[0,1\right),\;\phi\in\left[0,2\pi\right)\right\}.
\end{equation}
See an illustration of $\cA_{\mathsf{1D}}$ and its convex hull in Fig.~\ref{fig:momentCurve}~(c) and (d).
Writing $c_k = \left\vert c_k \right\vert e^{j\phi_k}$, $\bx$ can be represented as a positive combination of the atoms in $\cA_{\mathsf{1D}}$ as $\bx = \sum_{k=1}^r \left\vert c_{k} \right\vert e^{j\phi_k} \ba \left( \tau_k \right)$. It is easy to verify that $\cA_{\mathsf{1D}}$ is centrally symmetric around the origin, and consequently it induces an atomic norm over $\mathbb{C}^n$, as defined in \eqref{eq:atomic-norm-definition} and \eqref{eq:atomic-norm-decomposition}. It is worth noting that minimizing the atomic norm of $\bm{x}$ is equivalent to minimizing the total variation of $x(t)$, i.e.
\begin{equation}
\min \; \|x (t) \|_{\mathsf{TV}} \quad \mbox{s.t.}\quad \bx = \int_0^1 \ba(t) \mathrm{d}x(t),
\end{equation}
and both viewpoints are used frequently in the literature.

Remarkably, this atomic norm admits an equivalent semidefinite program (SDP) characterization, thanks to the Carath{\'e}odory--Fej{\'e}r--Pisarenko decomposition \cite{georgiou2007caratheodory}:
\begin{align}
\label{eq:atomic-norm-primal-SDP}
\left\Vert \bx\right\Vert_{\cA}=\inf_{\substack{\bu\in\mathbb{C}^{n}\\
t>0}}  \Big\{ \frac{1}{2n} \tr& \left(\toep\left(\bu\right)\right) +\frac{1}{2}t:\nonumber \\
& \begin{bmatrix}\toep\left(\bu\right) & \bx\\
\bx^{\mathsf{H}} & t
\end{bmatrix}\succeq0\Big\}.
\end{align}
Contrary to its abstract definition in \eqref{eq:atomic-norm-definition}, the reformulation \eqref{eq:atomic-norm-primal-SDP} provides a tractable approach to \emph{compute} the quantity $\anorm{\bx}$, which can be accomplished using generic off-the-shelf convex solvers \cite{grant2008cvx}. The Vandermonde decomposition of $\toep(\bu)$, i.e. $ \toep(\bu)= \sum_{k=1}^{r'} |c_k'| \ba(\tau_k')\ba(\tau_k')^{\sf{H}}$ can then be used to identify the support $\widehat{\mathcal{T}} = \{\tau_k'\}$ of the atomic representation of $\bx$, as well as the atomic norm $\|\bx\|_{\cA}=\sum_{k=1}^{r'} |c_k'|$.

\begin{textbox*}[t]
\begin{center}
{\bf From Bounded Polynomials to Linear Matrix Inequalities}\\
\end{center}

Many applications encountered in signal processing, systems and control theory involve comparing the magnitudes of two real trigonometric polynomials $R(\tau) = \Re \langle \ba(\tau), \br\rangle $ and $S(\tau ) = \Re \langle \ba(\tau), \bs \rangle $, e.g. bounding the frequency response of a finite impulse response filter by a desired shape. Although such inequalities, in appearance, require to be verified over a continuous set of parameters, they can easily be translated into linear matrix inequalities (LMI) of finite dimension, which are amenable to optimization.

Central to the equivalence is a {\em Gram parametrization} of real trigonometric polynomials \cite{lasserre2010moments,teke2017role,choi1995sums}, by noticing that every real trigonometric polynomial $R(\tau) = \Re \langle \ba(\tau), \br \rangle$ can be equivalently represented as a quadratic form $R(\tau) = \ba(\tau)^{\sf{H}}\bG\ba(\tau)$ for a family of Hermitian matrices $\bG\in\mathcal{G}(R)$, where $\bG$ is related to $\br$ through the Gram mapping:
\begin{equation}
\label{eq:GramRelation}
\bG\in\mathcal{G}(R) \Longleftrightarrow \tr(\bG)  = \Re(r_0), \; \sum_{i=1}^{n-k} G_{i,i+k} = \frac{r_k}{ 2},\;k=1,\dots,n-1.
\end{equation}
A remarkable property (e.g. \cite[Lemma 4.23]{dumitrescu2007positive}) is that the Gram mapping preserves the partial order between the polynomials and the Hermitian matrices: let $\bm{G}\in \mathcal{G}(R)$, then $R(\tau) \leq S(\tau)$ holds for every $\tau \in [0,1)$ if and only if there exists $\bm{H}\in \mathcal{G}(S)$ such that $\bm{G} \preceq \bm{H}$.
\begin{center}
    \includegraphics[width=0.5\textwidth]{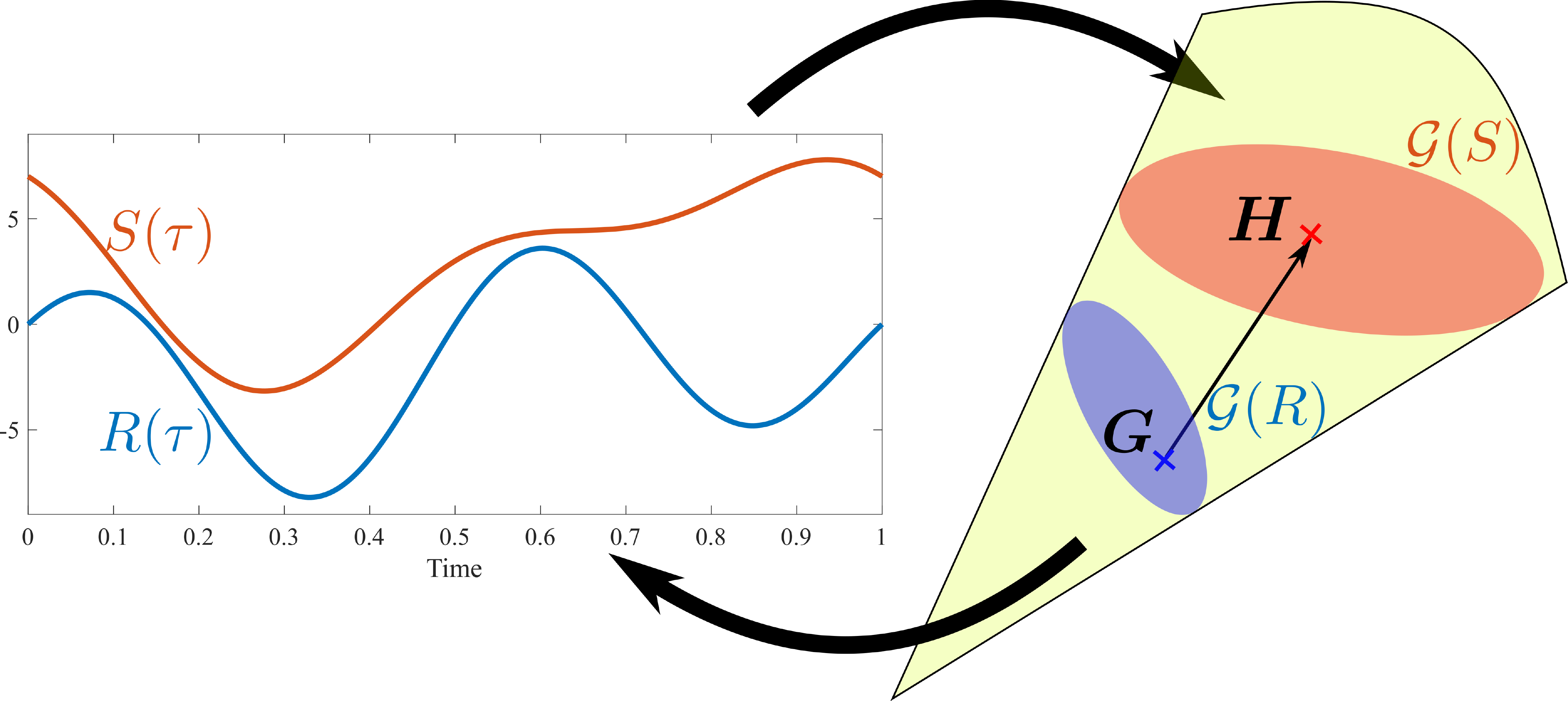}
    \captionof{figure}{Illustration of the Gram mapping for two polynomials satisfying $R(\tau) \leq S(\tau)$ over $[0,1)$, and the corresponding $\bm{G}$, $\bm{H}$ in the respective Gram sets such that $\bm{G} \preceq \bm{H}$.}
    \label{fig:GramMap}
\end{center}

As an example, consider the dual norm constraint $\anorm{\bp}^\ast \leq 1$ in \eqref{eq:abstract-dual}, which amounts to upper bounding $R(\tau)= |\langle \ba(\tau), \bp\rangle|^2$ by $S(\tau)=1$. Since $R(\tau) = \ba(\tau)^{\sf{H}}\bp \bp^{\sf{H}}\ba(\tau)$, it is clear that $\bm{p}\bm{p}^{\sf{H}} \in \mathcal{G}(R)$. The constraint holds if and only if their exists a matrix $\bH \in \mathcal{G}(S)$ satisfying
$ \bm{H} \succeq \bm{p}\bm{p}^{\sf{H}} .$
Rewriting this condition using the Schur's complement, as well as expanding the Gram mapping of $S(\tau)=1$, we can obtain the semidefinite constraint in \eqref{eq:1Ddual-SDP}, a consequence also known as the Bounded Real Lemma.

\end{textbox*}

\subsection{Duality and Atomic Decomposition}
\label{sub:DualityAndAtomicDecomposition}

The Lagrangian duality theory marks an important aspect in understanding the atomic norm. The Lagrange dual problem associated with the atomic norm minimization \eqref{eq:atomic-norm-definition} reads \cite{CandesFernandez2012SR}
\begin{equation}
\label{eq:abstract-dual}
\max_{\bp} \; \Re \left\langle \bx , \bp \right\rangle \quad \mbox{subject to} \; \anorm{\bp}^\ast \leq 1,
\end{equation}
where the dual atomic norm $\anorm{\bp}^{\ast}$ of a vector $\bp \in \mathbb{C}^n$ is defined with respect to the atomic set $\mathcal{A}_{\mathsf{1D}}$ as
\begin{align}\label{eq:dualAtomicNormDefinition}
    \anorm{\bp}^\ast  \triangleq & \;\, \sup_{\ba \in\cA_{\mathsf{1D}}} \Re \left\langle \ba ,\bp \right\rangle   = \sup_{\tau \in \left[0,1\right)} \vert \underbrace{\left\langle \ba \left( \tau \right),\bp \right\rangle}_{P(\tau)} \vert.
\end{align}
The last equality of \eqref{eq:dualAtomicNormDefinition} suggests that the dual atomic norm can be interpreted as the supremum of the modulus of a complex trigonometric polynomial $P(\tau) = \left\langle \ba \left( \tau \right), \bp \right\rangle = \sum_{\ell=0}^{n-1} p_\ell e^{-j2\pi \ell \tau } $ with coefficients given by the vector $\bp$. Constraints of this type is known to be equivalent to linear matrix inequalities involving the positivity of some Hermitian matrices (c.f. the box ``From Bound Polynomials to Linear Matrix Inequalities''). The dual program \eqref{eq:abstract-dual} can be reformulated into the SDP below:
\begin{align}
\label{eq:1Ddual-SDP}
\max_{\bp\in\mathbb{C}^n,\bH\in\mathbb{C}^{n\times n}} & \quad \Re \left\langle \bx,\bp\right\rangle \nonumber\\
\mbox{subject to } & \quad \begin{bmatrix}{\bm{H}} & \bp\\
\bp^{\mathsf{H}} & 1
\end{bmatrix}\succeq0\nonumber \\
& \quad \sum_{i=1}^{n-k} H_{i,i+k} = \delta_k ,\;  k=0,\ldots,n-1,
\end{align}
where $H_{i,j}$ is the $(i,j)$th entry of the matrix $\bH$, and the indicator function $\delta_k$ equals to $1$ if $k=0$ and $0$ otherwise.

\begin{figure}[t]
    \centering
     \includegraphics[width=0.5\textwidth]{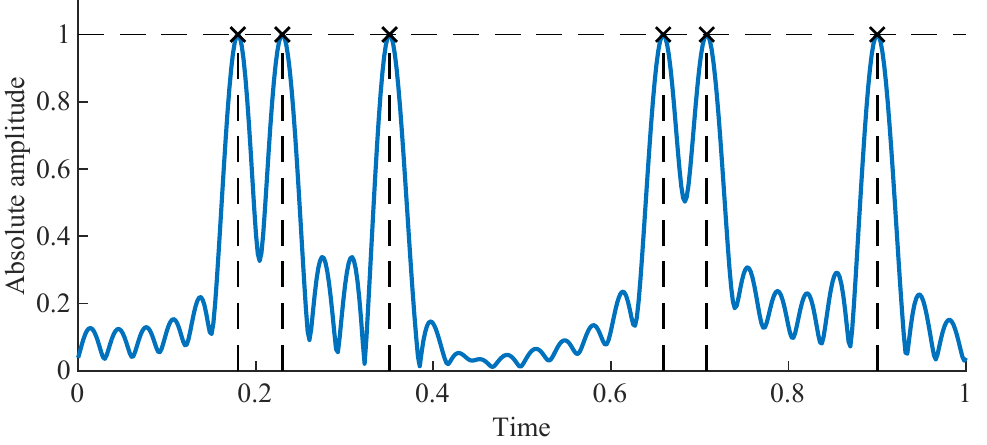}
    \caption{Spike localization via localizing the peaks of the dual polynomial $|\widehat{P}(\tau)|$ (in blue) associated with the optimal solution $\widehat{\bp}$ of the dual program \eqref{eq:abstract-dual} for a signal $\bx$ of length $n=33$ with $6$ true spikes (in black). }
    \label{fig:delayLocalisationFromDual}
\end{figure}

Another merit of the dual formulation is that the support set of the atomic decomposition can be inferred from the optimal solution $\widehat{\bp}$ to the dual problem \eqref{eq:abstract-dual}, by examining the dual polynomial $\widehat{P}(\tau) =  \left\langle \ba \left( \tau \right), \widehat{\bp} \right\rangle$. We identify the spikes as the locations of the extreme values of the modulus of $\widehat{P}(\tau)$:
\begin{equation}\label{eq:support_loc}
\widehat{\mathcal{T}} = \left\{\tau: |\widehat{P}(\tau)| = 1 \right\}.
\end{equation}
This is possible, because, under strong duality, both the primal and the dual problems must share the same optimal objective value, i.e.
$\anorm{\bx} = \sum_{k=1}^{r'} |c_k'|,$
where $\bx = \sum_{k=1}^{r'}c_k'\ba(\tau_k')$ is the {\em atomic} decomposition of $\bx$. Consequently, the optimal value of dual program becomes
\begin{align*}
 \Re \left\langle \bx ,\widehat{\bp} \right\rangle & =   \Re \left\langle  \sum_{k=1}^{r'}c_k'\ba(\tau_k') , \widehat{\bp} \right\rangle = \Re    \sum_{k=1}^{r'} {c_k'}^{*} \widehat{P}(\tau_k')  ,
\end{align*}
indicating $\widehat{P}(\tau_k') =\mbox{sgn}(c_k') = c_k'/|c_k'|$ whenever the atomic decomposition is non-vanishing at $\tau_k'$. This approach is illustrated in Fig.~\ref{fig:delayLocalisationFromDual} for a length-$33$ signal with 6 spikes, where the peaks of $\widehat{P}(\tau) $ matches with the locations of the true spikes, indicating the atomic decomposition perfectly recovers the true sparse representation.

The atomic norm offers an approach for line spectrum super resolution that is drastically different from traditional methods, which rely heavily on the correctness of model order estimation. The dual polynomial approach, in contrast, does not require any prior knowledge on the model order, and can estimate the spikes with an infinitesimal precision.

\subsection{Exact Recovery Guarantees} \label{sub:ExactRecovery}

So far, we have explained the algorithmic approach of atomic norm minimization for line spectrum super resolution. A central question is to understand whether this convex relaxation is tight or not. More precisely, once would like to identify the conditions under which the estimated support $\widehat{\mathcal{T}}$ coincides with the true support $\mathcal{T}$ of the signal $\bx$, and correspondingly, the atomic decomposition $\bx = \sum_{k=1}^{r'}c_k'\ba(\tau_k')$ coincides with the sparsest representation $\bx = \sum_{k=1}^r c_k \ba(\tau_k)$ over the atomic set~$\cA_{\mathsf{1D}}$.

Such questions were extensively addressed in the context of $\ell_1$ norm minimization, where the atomic set $\cA$ has a finite number of elements. The performance guarantees often depend on specific structural properties of $\cA$, formalized in the notion of restricted isometry property (RIP) \cite{candes2008restricted}, or certain incoherence properties \cite{TroppGilbert}. Unfortunately, these properties do not hold when considering a continuous dictionary such as $\cA_{\mathsf{1D}}$, since two atoms $\ba\left(\tau\right)$ and $\ba(\tau +\delta)$ can be more and more correlated with each other as their separation $\delta$ tends to zero, leading to arbitrarily small RIP or coherence constants.

Nonetheless, one could ask for which class of signals the relaxation is tight. Leveraging duality theory, the atomic norm approach is tight for a {\em fixed} signal $\bx$, i.e. $\widehat{\mathcal{T}} = \mathcal{T}$, as long as there exists a {\em dual certificate} $\bp_\star$, such that $P_\star (\tau)= \langle \ba(\tau), \bp_{\star}\rangle$ satisfies \cite{CandesFernandez2012SR}
	\begin{subequations}
		\begin{align}
			P_\star(\tau_k) &= \sgn \left( c_k \right), \quad \forall \tau_k \in \mathcal{T}, \label{eq:dualCertificate-interpolationCond}\\
			\left\vert P_\star(\tau) \right\vert &<1, \quad \forall \tau \notin \mathcal{T}.\label{eq:dualCertificate-extremalCond}
	\end{align}
	\end{subequations}
In other words, it amounts to find an $(n-1)$-order trigonometric polynomial that interpolates sign patterns of the spike signal at the spike locations, as well as is bounded in magnitude by 1.

\begin{figure}[t]
    \centering
    \includegraphics[width =  \columnwidth]{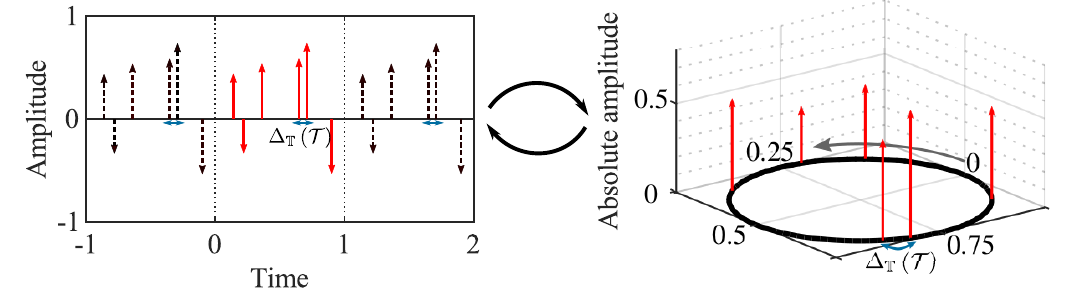}
    \caption{A representation of the minimal wrap-around distance $\Delta_{\mathbb{T}}\left(\mathcal{T}\right)$ for a give set of spikes $\mathcal{T}$. The distance corresponds to the minimal gap between any element  $\tau \in\mathcal{T}$ and any distinct elements in the \emph{aliased} set $\mathcal{T} + \mathbb{Z}$.  }
    \label{fig:WarpAroundDistance}
\end{figure}

Intuitively, the difficulty of interpolation depends on the separations between the spikes in $\mathcal{T}$, and more precisely on the \emph{minimal separation}, or the minimal wrap-around distance between any pair of distinct spikes in $\mathcal{T}$, defined formally as
\begin{equation}
\label{eq:warpAroundDistance}
\Delta_{\mathbb{T}}\left(\mathcal{T}\right)\triangleq\inf_{\substack{\tau,\tau^{\prime}\in \mathcal{\mathcal{T}}\\
\tau\neq \tau^{\prime}}}\min_{q\in\mathbb{Z}}\left|\tau-\tau^{\prime}+q\right|.
\end{equation}
This metric is illustrated in Fig.~\ref{fig:WarpAroundDistance}, and reflects the periodic behavior of the atom $\ba\left(\tau+q\right) = \ba\left(\tau\right)$ for any integer $q\in\mathbb{Z}$. For instance, if $\mathcal{T} = \{0.1, 0.9\}$, then $\Delta_{\mathbb{T}}\left(\mathcal{T}\right) = 0.2$.

A remarkable result, established by Cand\`es and Fernandez-Granda in \cite{CandesFernandez2012SR}, is that for sufficiently large $n$, a valid certificate can be constructed in a deterministic fashion, as long as the separation condition $\Delta_{\mathbb{T}}\left(\mathcal{T}\right)\geq \frac{4}{n-1}$
holds, regardless of the complex amplitudes of the spikes. Furthermore, this result does not make any randomness assumptions on the signal. Later, this separation condition has been further improved by Fernandez-Granda \cite{Fernandez-Granda2016sr} to
$$\Delta_{\mathbb{T}}\left(\mathcal{T}\right) > \frac{2.52}{n-1}.$$
Conversely, there exist some spike signals with $\Delta_{\mathbb{T}}\left(\mathcal{T}\right) < \frac{2}{n-1}$ such that atomic norm minimization fails to resolve \cite{Ferreira2018}.

\begin{textbox*}[t]
\begin{center}
{\bf  Is the Separation Condition Necessary?}\\
\end{center}

One might wonder if requiring a separation condition makes atomic norm minimization inferior, since many methods do not require such a separation in the noise-free case. However, some form of separation is unavoidable for {\em stable recovery} in noisy super resolution, no matter which method is used \cite{moitra2015super}. In particular, \cite{moitra2015super} shows that when $\Delta_{\mathbb{T}}(\mathcal{T}) < 2/n$, there exists a pair of spike signals $x(t)$ and $x^{\prime}(t)$ with the same minimal separation, such that no estimator can distinguish them. As an illustration, Fig.~\ref{fig:stabilityResolution} (a) exhibits such a pair of positive spike signals (c.f. \cite{moitra2015super} for its construction) with a minimal separation $1.7/n$, where their  observations are very close. Fig.~\ref{fig:stabilityResolution} (b) further demonstrates the distance between their observations as the signal dimension increases, for different separation parameter $\alpha$, where $\Delta_{\mathbb{T}}(\mathcal{T})=\alpha/n$. It is clear that their observations are increasingly indistinguishable as the signal dimension tends to infinity when $\Delta_{\mathbb{T}}(\mathcal{T}) < 2/n$.

 \begin{center}
 \begin{tabular}{cc}
    \includegraphics[width=0.45\textwidth]{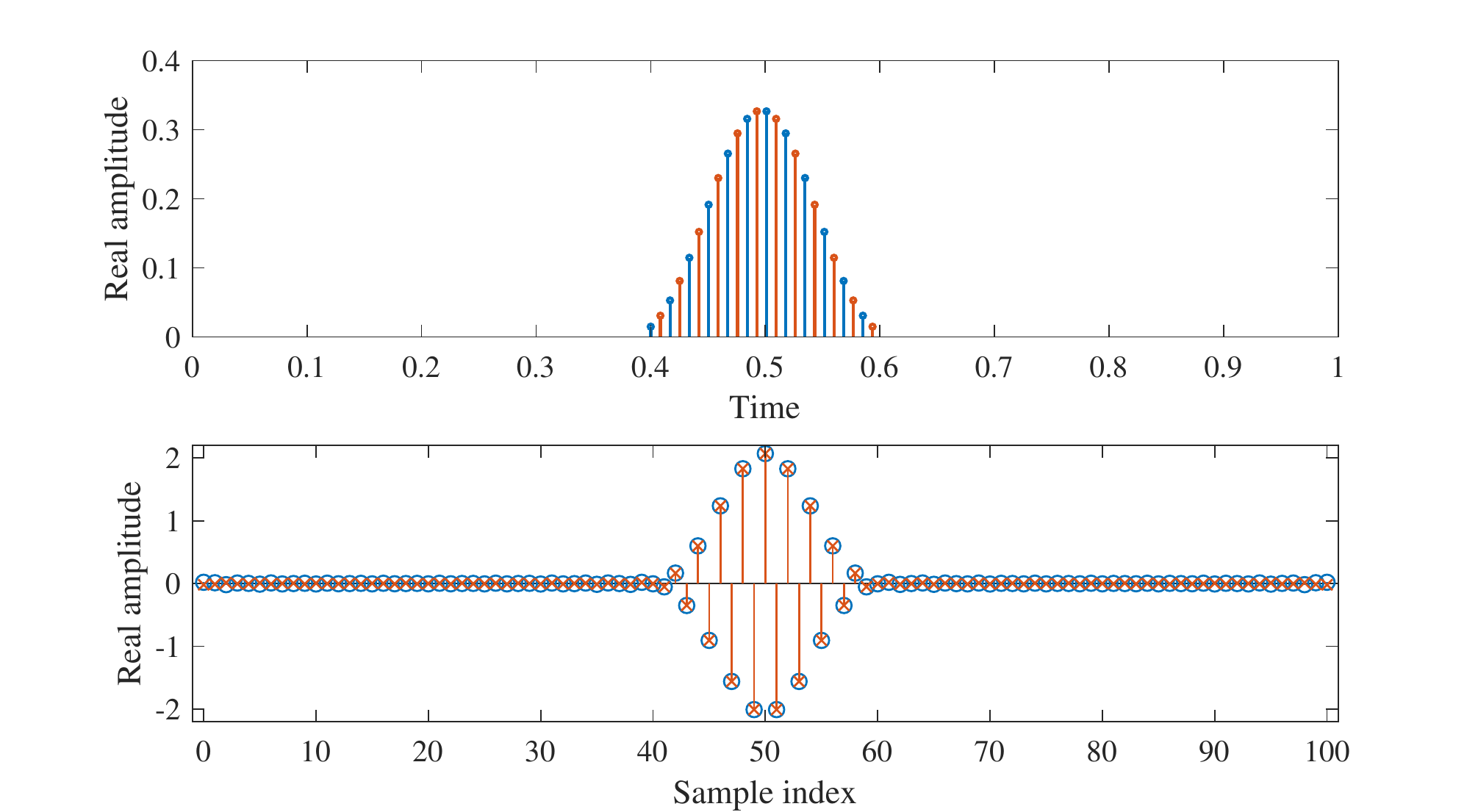} &  \includegraphics[width =  0.45\textwidth]{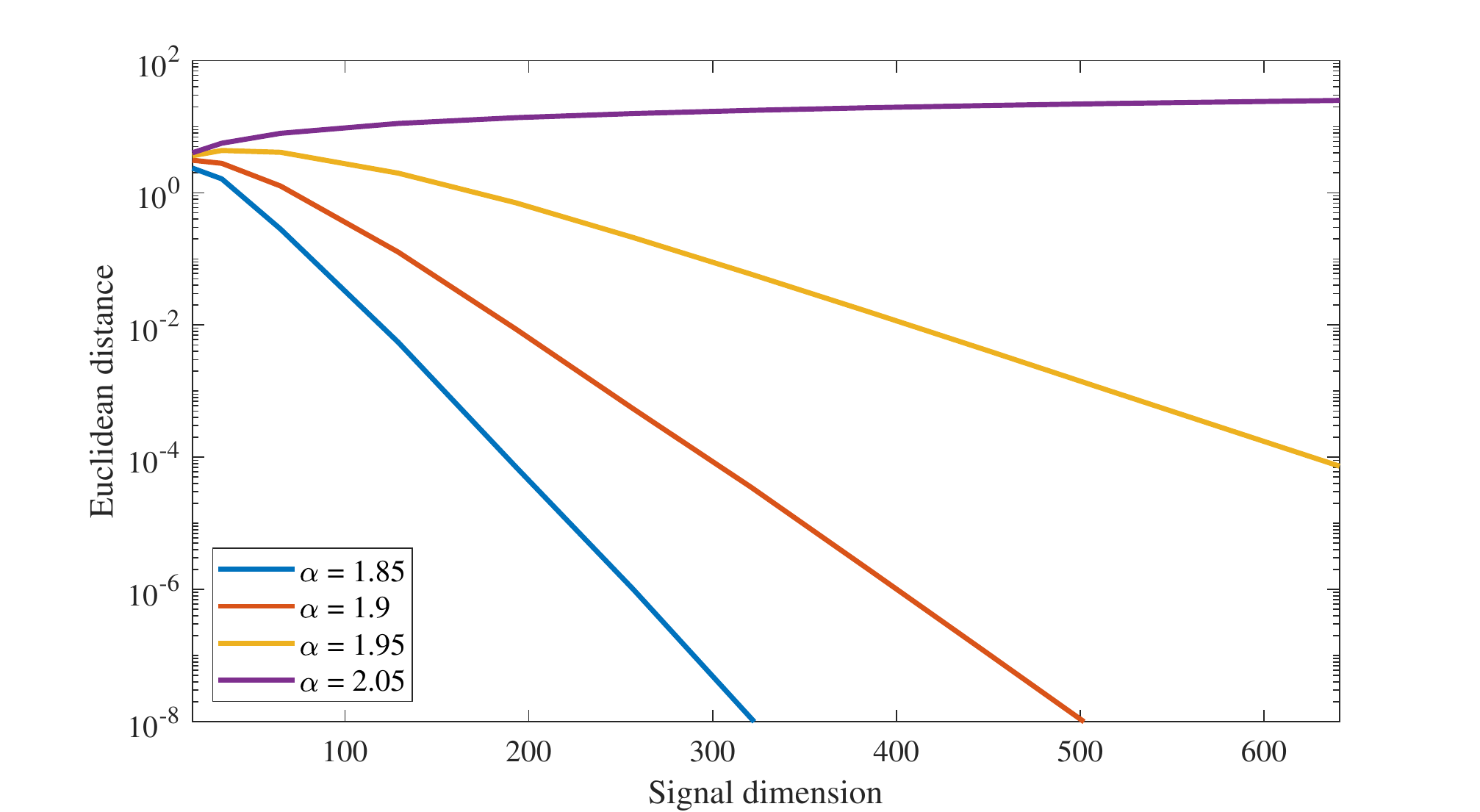} \\
    (a) & (b)
    \end{tabular}
    \captionof{figure}{(a) Illustration of a pair of spike signal $x(t)$ and $x^{\prime}(t)$ with a minimal separation $\Delta_{\mathbb{T}}(\mathcal{T}) = 1.7/n$ (top), yet their observations produced according to \eqref{eq:integral_form} are almost indistinguishable (bottom). (b) The Euclidean distance between the observations of $x(t),x^{\prime}(t)$ with a minimal distance $\Delta_{\mathbb{T}}(\mathcal{T})=\alpha/n$ as a function of the signal length $n$, for different values of the separation parameter~$\alpha$.}
        \label{fig:stabilityResolution}
\end{center}

\end{textbox*}

\subsection{Atomic Norm Denoising}\label{subsec:AtomicNormDenoising}

In practice, the observations are corrupted by noise, and no estimator can \emph{exactly} recover the spike signal $x(t)$. This raises a natural question regarding the \emph{robustness} of the estimate produced by atomic norm minimization methods. When the noise is additive and the observation $\bz$ obeys the noisy model \eqref{eq:model_withoutPSF}, it has been proposed to estimate $\bx$ by searching around the observation $\bz$ for signals with small atomic norms~\cite{bhaskar2013atomic}:
\begin{equation}\label{eq:DenoisingAlgorithm}
\min_{\bx} \quad \frac{1}{2} \left\Vert  \bx - \bz  \right\Vert_2^2 + \lambda \anorm{\bx},
\end{equation}
where $\lambda > 0$ is a regularization parameter that draws a trade-off between the fidelity to the observation and the size of the atomic norm. This method, known as ``atomic norm denoising'', can be interpreted as a generalization of the celebrated LASSO estimator \cite{Tib96}.

When the noise vector $\widetilde{\bm{\epsilon}}$ is composed of i.i.d. complex Gaussian entries with zero mean and variance $\sigma^2$, the mean squared error (MSE) of the estimate $\widehat{\bx}$ returned by \eqref{eq:DenoisingAlgorithm} can be bounded as \cite{bhaskar2013atomic}
\begin{equation*}
    \frac{1}{n}  \left\Vert \widehat{\bx} - \bx  \right\Vert_2^2 = O\left( \sigma \sqrt{\frac{\log n}{n}} \sum_{k=1}^r | c_k | \right)
\end{equation*}
with high probability by setting $\lambda = \eta\sigma\sqrt{n\log n}$ for some constant $\eta\in(1,\infty)$, e.g. $\eta=1.2$ in practice. This error rate can be significantly improved when the spikes satisfy the separation condition $\Delta_{\mathbb{T}}\left(\mathcal{T}\right)\geq \frac{4}{n-1}$, where with high probability one has \cite{tang2015near}
\begin{equation*}
    \frac{1}{n}  \left\Vert \widehat{\bx} - \bx  \right\Vert_2^2 = O\left( \sigma^2 \frac{r \log n}{n} \right).
\end{equation*}
This last error rate is near-optimal up to some logarithmic factor, since no estimator can achieve an MSE below the rate $O\left(\sigma^2\frac{r\log\left(n /r \right)}{n} \right)$ \cite{tang2015near}.

A more important performance criteria in super resolution concerns the stability of the support estimate $\widehat{\mathcal{T}}$, which has been studied in \cite{tang2015near}\nocite{fernandez2013support,li2018approximate,Duval2015exact}--\cite{da2019stable}. When the spikes satisfy the separation condition $\Delta_{\mathbb{T}}\left(\mathcal{T}\right) > \frac{5.0018}{n-1}$, and the complex amplitudes of the coefficients $\left\{c_k \right\}_{1\leq k\leq r}$ have approximately the same modulus, then it is established in \cite{li2018approximate} that the atomic decomposition of the output $\widehat{\bx}$ of \eqref{eq:DenoisingAlgorithm} is  composed of the same number of spikes, i.e. $|\widehat{\mathcal{T}}| = |\mathcal{T}|=r$ and that the estimated parameters satisfy
\begin{align*}
    \left\vert c_k \right\vert \left\vert \widehat{\tau}_k - \tau_k \right\vert &= O\left(\sigma \frac{\sqrt{\log n}}{n^{3/2}}  \right), \;
    \left\vert c_k - \widehat{c}_k\right\vert = O\left(\sigma \sqrt{\frac{\log n}{n}}  \right)
\end{align*}
with high probability. Altogether, it can be seen that atomic norm denoising achieves near-optimal performance guarantees as long as the spikes are separated by a few times the Rayleigh limit (c.f. the box ``Is the Separation Condition Necessary?'').

\begin{figure}[ht]
    \centering
    \includegraphics[width=0.49\textwidth]{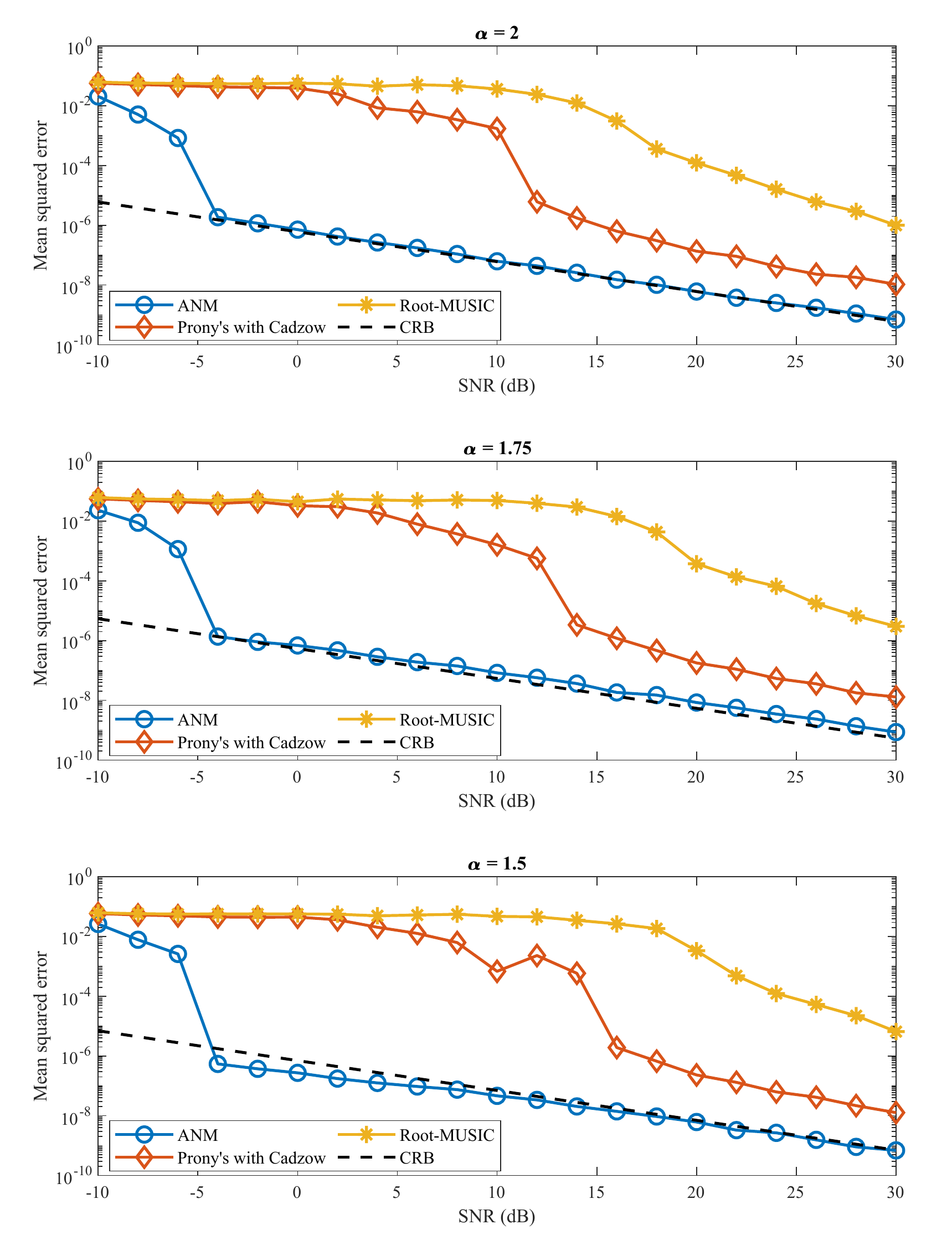}
      \caption{The MSE of different methods for estimating two spikes with opposite signs separated by $\Delta_{\mathbb{T}}(\mathcal{T})=\alpha/n$, averaged over $200$ Monte Carlo trials and benchmarked against the CRB for different separation parameter $\alpha=2$, $1.75$, $1.5$. Here, the signal length is $n=101$.}
    \label{fig:performanceComparison_negSpikes}
\end{figure}

It is natural to wonder how atomic norm denoising fares compared with classical approaches such as Prony and MUSIC for line spectrum estimation. We examine their ability to resolve close-located spikes with {\em opposite} signs, where the reconstruction performance is measured in terms of the MSE of the estimated spike locations $\widehat{\mathcal{T}}$, and for different values of separation  $\Delta_{\mathbb{T}}(\mathcal{T}) = \alpha /n$. Fig.~\ref{fig:performanceComparison_negSpikes} shows the MSE of atomic norm denoising, Prony's method with Cadzow denoising \cite{Blu2008Sparse}, and root-MUSIC \cite{rao1989performance} with respect to the SNR defined as $\|\bx\|_2^2/(n\sigma^2)$,  benchmarked against the Cram\'er-Rao bound (CRB), when the separation parameter $\alpha = 2$, $1.75$, $1.5$ respectively. It is clear that atomic norm denoising outperforms classical approaches, and approaches the CRB  at a much lower SNR.

\subsection{A Faster Algorithm via ADMM}

While the SDP formulation is tractable, its computational complexity is prohibitive when solving large-dimensional problems. Fortunately, it is possible to develop tailored algorithms that are significantly faster. For conciseness, we will discuss one approach based on the Alternating Direction Method of Multipliers (ADMM) \cite{bhaskar2013atomic}. The general principle of ADMM is to split the quadratically-augmented Lagrangian function of an optimization problem into a sum of separable sub-functions \cite{Boyd2010ADMM}. Each iteration of the algorithm consists of performing independent local minimization on each of those quantities, while ensuring that the feasibility constraints are always satisfied. The iterations run until both primal and dual residuals satisfy a pre-defined tolerance level.

We take atomic norm denoising \eqref{eq:DenoisingAlgorithm} as an example, which, in light of \eqref{eq:atomic-norm-primal-SDP}, can be equivalently rewritten as
\begin{align*}
\min_{\boldsymbol{x},\boldsymbol{u}, t}&\quad  \frac{1}{2}\left\Vert\bm{x} - \bm{z} \right\Vert_2^{2} + \frac{\lambda}{2}\left(\frac{1}{n}\mbox{Tr}\left(\toep\left(\boldsymbol{u}\right)\right)+t\right)\\
 \mbox{subject to} & \quad  \boldsymbol{S}=\begin{bmatrix}
\toep\left(\boldsymbol{u}\right) & \boldsymbol{x} \\
\boldsymbol{x}^{\mathsf{H}} & t \end{bmatrix},\;\boldsymbol{S}\succeq 0.
\end{align*}
The above program has been ``augmented'' by introducing an intermediate variable $\bm{S}$ for the purpose of decoupling the positive semidefinite constraint on the matrix $\bm{S}$ from the linear constraints on its structure. The augmented Lagrangian $\mathcal{L}$ is given as
\begin{align*}
\mathcal{L}\left(\boldsymbol{x},\boldsymbol{u},t,\boldsymbol{\Sigma},\boldsymbol{S}\right)= & \frac{1}{2}\left\Vert \bm{x} - \bm{z}  \right\Vert_2^{2} + \frac{\lambda}{2}\left(\frac{1}{n}\mbox{Tr}\left(\toep\left(\boldsymbol{u}\right)\right)+t\right)\\
&+\left\langle \boldsymbol{\Sigma},\boldsymbol{S}-\begin{bmatrix}
\toep\left(\boldsymbol{u}\right) & \boldsymbol{x} \\
\boldsymbol{x}^{\mathsf{H}} & t \end{bmatrix} \right\rangle \\
&+\frac{\rho}{2}\left\Vert\boldsymbol{S}-\begin{bmatrix}
\toep\left(\boldsymbol{u}\right) & \boldsymbol{x} \\
\boldsymbol{x}^{\mathsf{H}} & t \end{bmatrix}\right\Vert^{2}_{\mathrm{F}},
\end{align*}
where $\boldsymbol{S}$ and $\boldsymbol{\Sigma}$ are $(n+1)$-dimensional Hermitian matrices, and $\rho>0$ is a regularization parameter. The successive update steps to minimize the augmented Lagrangian are given in Alg.~\ref{alg:ADMM}. Closed-form solutions can be found for the first update step, yielding a very efficient implementation. The second update is the most costly part, as a projection over the cone of positive semidefinite Hermitian matrices has to be computed. This computation is typically achieved using power methods~\cite{Golub2000Eigenvalue}, with a computational complexity of $O(n^3)$  per iteration.

 \renewcommand{\algorithmicrequire}{\textbf{Input:}}
\renewcommand{\algorithmicensure}{\textbf{Output:}}
\begin{algorithm}[ht]
	\caption{ADMM for atomic norm denoising \cite{bhaskar2013atomic}}
	\label{alg:ADMM}
	\begin{algorithmic}
	\REQUIRE Observation $\bz$; Parameters $\lambda,\rho>0$; \\
	\STATE Initialize $ j=0$, and $\bm{\Sigma}_0, \bm{S}_0$ to zero matrices\\
	\textbf{repeat} until stopping criteria
		\begin{align*}
\;&\left(\boldsymbol{x}_{j+1},\boldsymbol{u}_{j+1},t_{j+1}\right) \gets \argmin_{\boldsymbol{x},\boldsymbol{u},t} \mathcal{L} \left(\boldsymbol{x},\boldsymbol{u},t,\boldsymbol{\Sigma}_{j},\boldsymbol{S}_{j}\right);\\
\;&\boldsymbol{S}_{j+1}\gets\argmin_{\boldsymbol{S}\succeq 0} \mathcal{L}\left(\boldsymbol{x}_{j+1},\boldsymbol{u}_{j+1},t_{j+1},\boldsymbol{\Sigma}_{j},\boldsymbol{S}\right);\\
\;&\boldsymbol{\Sigma}_{j+1}\gets \boldsymbol{\Sigma}_{j}+\rho\left(\boldsymbol{S}_{j+1}-\begin{bmatrix}
\toep\left(\boldsymbol{u}_{j+1}\right) & \boldsymbol{x}_{j+1} \\
\boldsymbol{x}_{j+1}^{\mathsf{H}} & t_{j+1} \end{bmatrix}\right);\\
\;& j\gets j+1;
\end{align*}
	\ENSURE $\bm{x}_j$
	\end{algorithmic}
\end{algorithm}

\subsection{Can we discretize?}

It may be worthwhile to pause and compare atomic norm minimization to other approaches based on convex optimization for super resolution, in particular, $\ell_1$ minimization that is widely popular for high-resolution imaging and localization in the recent literature due to Compressed Sensing (CS) \cite{CandesTao2006,Donoho2006}.

The $\ell_1$ norm can be seen as a discrete approximation of the atomic norm. Indeed, taking the atomic set $\cA_{\mathsf{1D}}$, one can pick a desired resolution $Q$ and discretize it as:
$$ \cA_{\mathsf{1D, discrete}} =\left\{ e^{j\phi} \ba\left( \frac{q}{Q} \right) :   q=0,\ldots, Q-1 ,\;\phi\in\left[0,2\pi\right)\right\},$$
and then perform $\ell_1$ minimization over $\cA_{\mathsf{1D,discrete}}$. The convex hull of $\cA_{\mathsf{1D, discrete}}$ approaches that of $\cA_{\mathsf{1D}}$ as the discretization gets finer, which suggests the performance of $\ell_1$ minimization over the discretized dictionary approaches that of atomic norm minimization asymptotically \cite{Tang2013sparserecovery}. If the spike signal meets a so-called ``non-degenerate source condition'' \cite[Definition 2]{duval2017sparse_partI}, this approach will return a sparse solution supported on the elements of the discretized $\cA_{\mathsf{1D,discrete}}$ surrounding the ground truth spikes, when the noise is small enough \cite{duval2017sparse_partI,duval2017sparse_partII}.\footnote{However, it remains unclear which class of spike signals satisfies the non-degenerate source condition in practice.}

However, this discretization may come with several undesired consequences when the grid size $Q$ is finite in practice. The theory of $\ell_1$ minimization only provides exact recovery guarantees when the spikes of $x(t)$ lie on the grid, which is unrealistic. In fact, there is always an inevitable {basis mismatch} \cite{Chi2011sensitivity}, between the spikes represented in the discretized dictionary $\cA_{\mathsf{1D,discrete}}$ and the true spikes, no matter how fine the grid is. Perfect recovery is not possible in this situation even in the absence of noise due to this mismatch. Furthermore, one can find signals whose representations in $\cA_{\mathsf{1D,discrete}}$ are not compressible due to spectral leakage, and therefore are poorly recovered using $\ell_1$ minimization, e.g. the recovery may contain many spurious spikes. Therefore, cautions are needed to account for such consequences when applying discretization, and efforts to mitigate the basis mismatch have been proposed extensively, e.g. \cite{tan2014joint,mamandipoor2016newtonized}.

\section{Generalizations of Atomic Sets}
\label{sec:other_atomic_sets}

The tool of atomic norms can be extended easily to handle a wide range of scenarios in a unified manner, by properly adjusting the atomic set for signal decompositions, such as incorporating prior information, dealing with multi-dimensional settings and multiple measurement vectors, to illustrate a few.

\subsection{Atomic Set for Positive Spikes}

In some applications, there exist additional information about the spikes, such as the coefficients of the spikes in \eqref{eq:single_mv} are positive, i.e. $c_k >0$. Examples include neural spike sorting, fluorescence microscopy imaging, or covariance-based spectrum estimation for noncoherent sources \cite{stoica1997introduction}. 

In this case, the atomic set reduces to the  moment curve $\cA_{0}$ in \eqref{eq:moment_curve}. The induced $\|\bx\|_{\cA}$ is no longer a norm, since $\cA_0$ is not centrally symmetric, but nonetheless, similar SDP characterization still holds. To be specific, the dual program now becomes
\begin{equation*}
\max_{\bp\in\mathbb{C}^n} \; \Re \left\langle \bx , \bp \right\rangle \quad \mbox{subject to} \; \sup_{\tau \in [0,1)}   \Re \left\langle \ba(\tau) ,\bp \right\rangle \leq 1,
\end{equation*}
where the constraint bounds the real part of the trigonometric polynomial $P(\tau) = \left\langle \ba \left( \tau \right), \bp \right\rangle$. Using the Fej\'er-Riesz Theorem (see e.g. \cite[Theorem 1.1]{dumitrescu2007positive} and the box ``From Bounded Polynomials to Linear Matrix Inequalities''), this can be equivalently represented as:
\begin{align*} 
\max_{\bp\in\mathbb{C}^n, \bH\succeq0 } & \;  \Re \left\langle \bx,\bp\right\rangle\\
\mbox{subject to } & \;  
\tr(\bH) + \Re (p_0) = 1,   \nonumber \\
&\; \sum_{i=1}^{n-k} H_{i,i+k} + p_k /2  = 0, \; k=1,\ldots, n-1. 
\end{align*}

\begin{figure}[ht]
    \centering
 \includegraphics[width=0.49\textwidth]{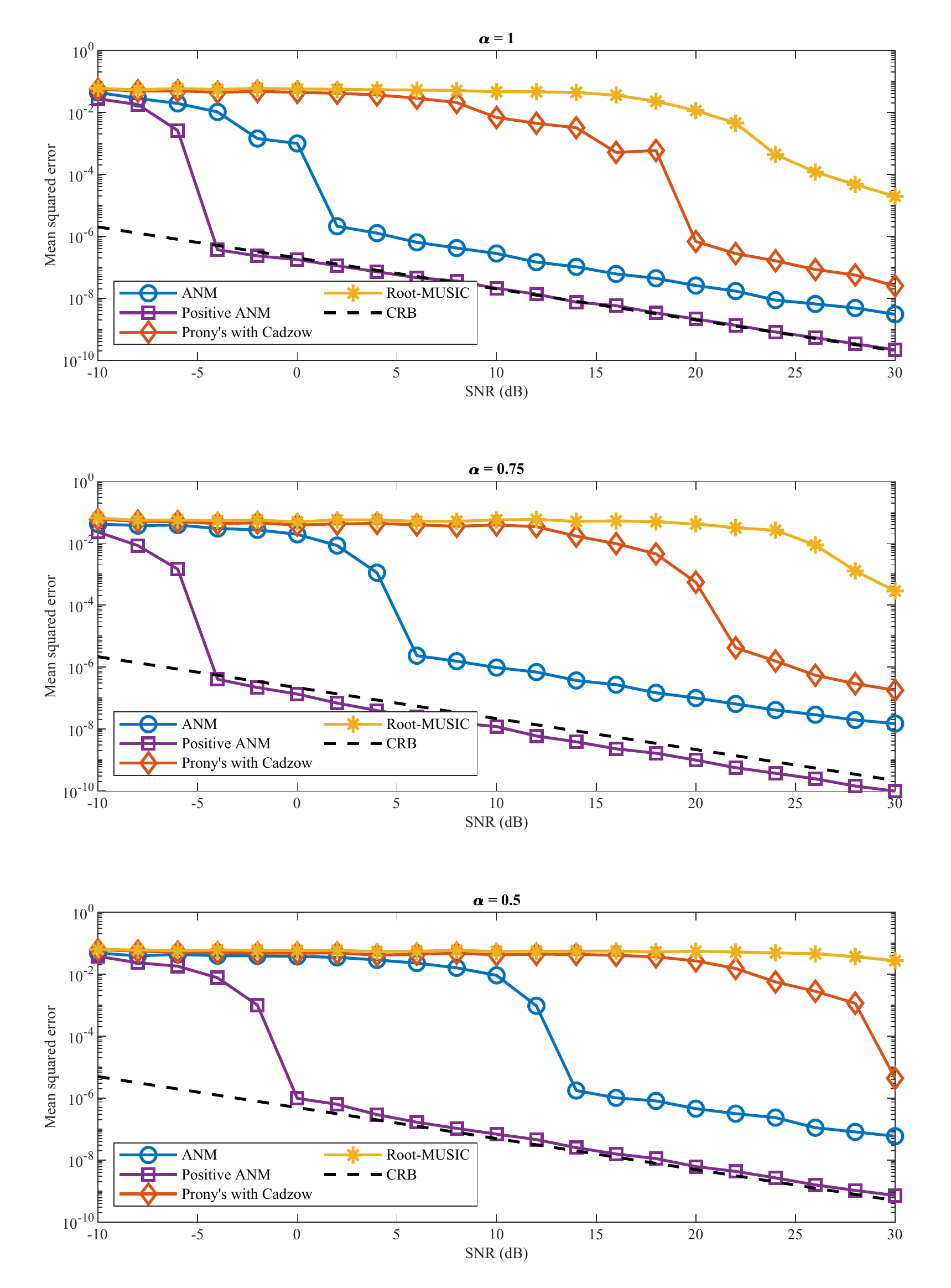}   
    \caption{The MSE of different methods for estimating two spikes with positive signs separated by $\Delta_{\mathbb{T}}(\mathcal{T})=\alpha/n$, averaged over 200 Monte Carlo trials and benchmarked against the CRB for different separation parameter $\alpha=1$, $0.75$, $0.5$. Here, the signal length is $n=101$.}
    \label{fig:performanceComparison_posSpikes}
\end{figure}

It is long established \cite{donoho2005sparse,fuchs2005sparsity} that the spikes can be perfectly localized as long as $r\leq \lfloor (n-1)/2 \rfloor$, {\em without} requiring any separation between the spikes, as long as they are positive. The stability of this approach as well as the implications of nonnegative constraints for other atomic sets are further studied in \cite{denoyelle2017support,schiebinger2017superresolution,eftekhari2018sparse,morgenshtern2016super}. As a comparison, Fig.~\ref{fig:performanceComparison_posSpikes} shows the MSE of atomic norm denoising with and without positive constraints, Prony's method with Cadzow denoising \cite{Blu2008Sparse}, and root-MUSIC \cite{rao1989performance} with respect to the SNR defined as $\|\bx\|_2^2/(n\sigma^2)$ for resolving two spikes with {\em positive} signs, separated by $\Delta_{\mathbb{T}}(\mathcal{T}) = \alpha/n$ for $\alpha = 1, 0.75, 0.5$ respectively. It can be seen that atomic norm denoising still outperforms classical approaches, and in particular, incorporating the positive constraint leads to further improvements.
 
\subsection{Atomic Set for Multi-Dimensional Spikes}

\begin{figure}[t]
    \centering
    \includegraphics[width=0.49\textwidth]{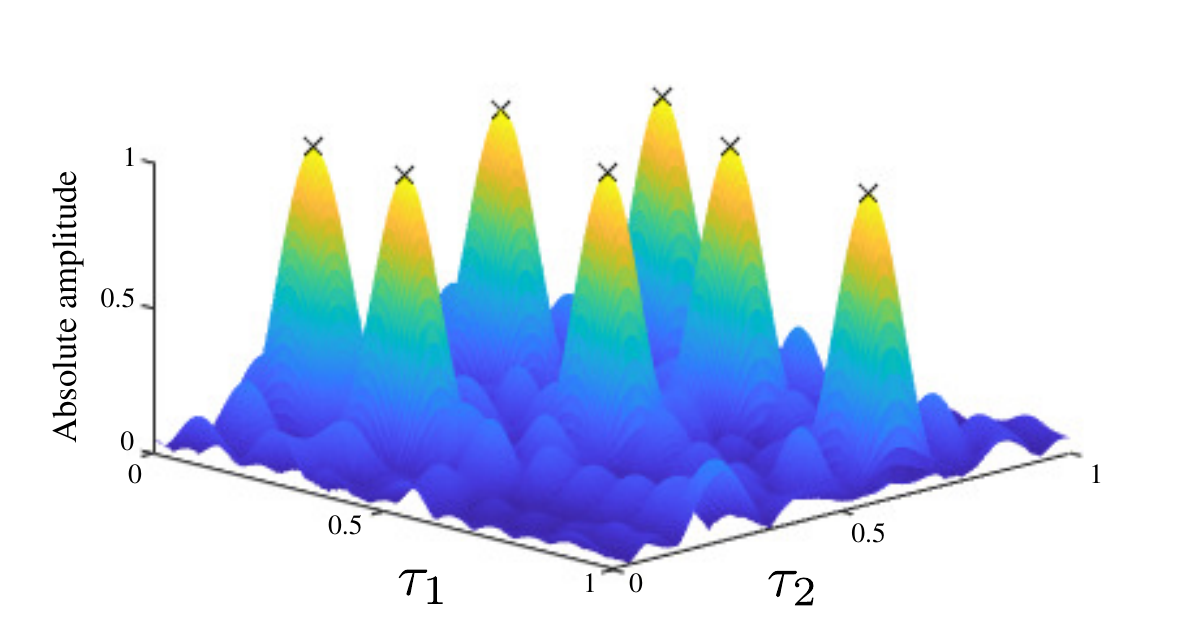}
    \caption{An illustration of spike localization using the dual polynomial approach for two-dimensional spikes via atomic norm minimization. Here, we set $n_1=12$, $n_2=10$ and $r=7$.}
    \label{fig:2Ddual}
\end{figure}

When the spikes reside in a multi-dimensional space, one can extend the one-dimensional model in a straightforward manner. Here, we illustrate the setup for the two-dimensional case, where 
each entry of the signal $\boldsymbol{X}_{\mathsf{2D}}\in\mathbb{C}^{n_1\times n_2}$ can be expressed as a superposition of $r$ complex sinusoids propagating in two directions:
\begin{equation}
\label{eq:2DatomicModel}
\boldsymbol{X}_{\mathsf{2D}} = \sum_{k=1}^r c_k \boldsymbol{a}_1(\tau_{1,k}) \boldsymbol{a}_2(\tau_{2,k})^{\top}, 
\end{equation}
where $c_k$ and $\bm{\tau}_k = \left[\tau_{1,k},\tau_{2,k}\right]^{\mathsf{T}} \in [0,1)^2$ are the complex amplitude and location of the $k$th spike, and $\boldsymbol{a}_i(\tau)$ is given by \eqref{eq:spectral_atom} with the dimension parameter replaced by $n_i$, $i=1,2$. It is natural to define the corresponding atomic set as \cite{CandesFernandez2012SR,chi2015compressive}
\begin{equation*} 
\mathcal{A}_{\mathsf{2D}} = \left\{ e^{j\phi}\ba_1 ( \tau_1 )\ba_2 ( \tau_2 )^{\top} : \; \bm{\tau} \in \left[0,1\right)^2,\;\phi\in\left[0,2\pi\right)\right\},
\end{equation*}
and the atomic norm according to \eqref{eq:atomic-norm-definition}. To localize the spikes, one could similarly study the associated dual problem,
\begin{align*}
\max_{\bP\in\mathbb{C}^{n_1\times n_2}} \; \Re \left\langle \bX, \bP  \right\rangle \quad \mbox{subject to} \; \anorm{\bP}^\ast \leq 1,
\end{align*}
where the dual atomic norm can be reinterpreted as the supremum of a bivariate complex trigonometric polynomial $P\left(\tau_1,\tau_2\right) = \langle  \ba_1 ( \tau_1 )\ba_2 ( \tau_2 )^{\top}, \bP \rangle$ with the matrix $\bP$ as its coefficients. Again, one can localize the spikes by examining the extremal points of the dual polynomial, which is illustrated in Fig.~\ref{fig:2Ddual}. Cautions need to be taken when attempting to solve the dual program in two or higher dimensions, since the Bounded Real Lemma \cite{lasserre2010moments,dumitrescu2007positive} does not hold anymore. Instead, a precise characterization requires solving a hierarchy of sum-of-squares relaxations, and fortunately in practice, the first level usually suffices \cite{xu2014precise,lasserre2010moments,chi2015compressive}. 

The tightness of the atomic norm minimization approach is closely related to a separation condition analogous to the one-dimensional case \cite{Fernandez-Granda2016sr}. Namely, the atomic decomposition is unique and exact, as soon as there exists a universal constant $C>0$ such that the set of spikes $\mathcal{T}=\{ \bm{\tau}_k\}_{1\leq k\leq r}$ satisfies
\begin{equation*} 
\Delta_{{\mathbb{T}}^2}\left(\mathcal{T}\right)\triangleq\inf_{\substack{\bm{\tau},\bm{\tau}^{\prime}\in \mathcal{\mathcal{T}}\\
\bm{\tau}\neq \bm{\tau}^{\prime}}}\min_{\bm{q}\in\mathbb{Z}^2}\left\Vert \bm{\tau}- \bm{\tau}^{\prime}+\bm{q}\right\Vert_\infty > \frac{C}{\min \left(n_1,n_2\right)-1}.
\end{equation*}
Moreover, if the signal $\bm{X}_{\mathsf{2D}}$ is real-valued, $C=4.76$ suffices to guarantee exact recovery of the spikes.

\subsection{Atomic Set for Multiple Measurement Vectors}

One can collect multiple snapshots of observations, where they share the same spike locations with varying coefficients. Consider $T$ snapshots, stacked in a matrix, $\bX_{\mathsf{MMV}}=[\bx_1,\ldots,\bx_T]$, which is expressed similarly to \eqref{eq:single_mv} as
\begin{equation}\label{eq:mmv_model}
\bX_{\mathsf{MMV}} = \sum_{k=1}^r  \ba(\tau_k) \bm{c}_k^{\top} ,
\end{equation}
where $\bc_k = [c_{1,k},\ldots,c_{L,k}]\in\mathbb{C}^T$ is the coefficient of the $k$th spike across the snapshots. Following the recipe of atomic norms, we define the atoms as
$$ \bA(\tau,\bb ) = \ba(\tau)\bb^\top , $$
where $ \tau \in [0,1)$, $\bb\in\mathbb{C}^{T}$ with $\|\bb\|_2 =1$. The atomic set is defined as
$$ \cA_{\mathsf{MMV}} = \left\{\bA(\tau,\bb ): \; \tau \in [0,1), \; \|\bb \|_2 =1 \right\}.$$

The atomic norm can be then defined following \eqref{eq:atomic-norm-definition} which turns out sharing similar nice SDP characterizations for primal and dual formulations as for the single snapshot model \cite{li2016offgrid}. The atomic norm $\|\bX_{\mathsf{MMV}} \|_\cA$ can be written equivalently as
\begin{align} \label{eq:atomic_norm_mmv}
 \| \bX_{\mathsf{MMV}} \|_\cA = \inf_{\substack{\bu\in\mathbb{C}^n \\ \bW\in\mathbb{C}^{T\times T}}} & \Big\{ \frac{1}{2n}\tr(\toep(\bu)) + \frac{1}{2}\tr(\bW) : \nonumber \\
 & \begin{bmatrix}
\toep(\bu) & \bX_{\mathsf{MMV}}  \\
\bX_{\mathsf{MMV}}^{\mathsf{H}} & \bW \end{bmatrix} \succeq 0 \Big\}.
\end{align}
A curious comparison can be drawn to the nuclear norm by noticing that one recovers the nuclear norm of $\bX_{\mathsf{MMV}} $ by replacing the principal block $\toep(\bu)$ in \eqref{eq:atomic_norm_mmv} with an arbitrary positive semidefinite matrix. The fact that $\toep(\bu)$ has significantly fewer degrees of freedom ($n$ versus $n^2$) is in parallel to that $\ba(\tau)$ has significantly fewer degrees of freedom than an arbitrary vector ($1$ versus $n$).

\begin{figure}[t!]
    \centering
    \includegraphics[width=0.47\textwidth]{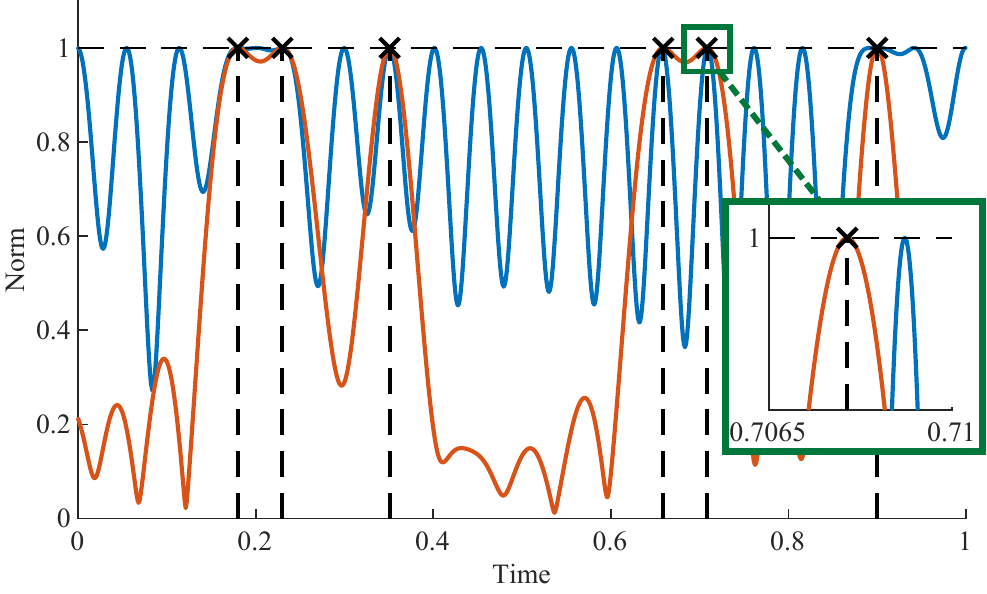}
    \caption{The dual polynomial using multiple snapshots (in red) successfully localizes all the spikes while the one using a single snapshot (in blue) fails for the same spike signal Here, $r=6$, $n=21$, and  $T=6$. }
    \label{fig:MMVpolynomial}
\end{figure}

Again, one can determine the atomic decomposition and localize the spikes by resorting to the dual program in a similar fashion. Fig.~\ref{fig:MMVpolynomial} illustrates an example that multiple snapshots improve the performance of localization over the single snapshot case when the coefficients across snapshots exhibit some kind of diversity, e.g. generated with i.i.d. complex Gaussian entries. 

\section{Generalizations of Measurement Models} 
\label{sec:beyond_denoising}

So far, we have seen that atomic norm minimization provides a means for super resolution via convex relaxation in additive Gaussian noise. The framework of convex optimization is quite versatile and can be extended to handle models when the measurements are partially observed, corrupted, contain interfering sources, or even come from unknown modulations. This is an important advantage over classical methods such as MUSIC or ESPRIT, as most of them cannot be extended easily to these variants of models.   

\subsection{Compressed spectral sensing}

CS \cite{CandesTao2006,Donoho2006} has suggested that it is possible to recover a signal using a number of measurements that is proportional to its degrees of freedom, rather than its ambient dimension. Consider the problem where only a subset of entries of $\bm{x}$ is observed,
\begin{equation*}
\bm{y}_{\mathsf{CS}} = \bm{A}_{\mathsf{CS}} \bm{x},
\end{equation*}
where $\bm{A}_{\mathsf{CS}} \in\mathbb{C}^{m\times n}$, and $m\ll n$ representing compressive acquisition of the signal $\bx$. The goal is to recover $\bm{x}$ and its spectral content from $\bm{y}_{\mathsf{CS}}\in\mathbb{C}^m$, the compressive measurements. This has applications in wideband spectrum sensing and cognitive radio \cite{tian2007compressed}, for example.  

One can easily extend the framework of atomic norm minimization, and recover $\bm{x}$ by solving the following program
\begin{equation*}
    \min_\bx \; \anorm{\bx} \;\;\; \mbox{ subject to } \;\;\;  \by_{\mathsf{CS}} = \bA_{\mathsf{CS}} \bx.
\end{equation*}
When $\bm{A}_{\mathsf{CS}}$ is a partial observation matrix, namely, a subset of $m$ entries of $\bx$ is observed uniformly at random, then $\bx$ can be perfectly recovered with high probability using $m = O \left( \log^2 n +  r \log r \log n \right)$ measurements as long as $\bx$ satisfies the separation condition, and with random signs of the coefficients $c_k$'s \cite[Theorem II.3]{tang2012compressed}. More generally, a broader class of measurement matrices $\bA_{\mathsf{CS}}$ can be allowed where its rows are drawn independently from some isotropic and incoherent distribution \cite{candes2011probabilistic,heckel2018generalized}, and exact recovery is possible under the same separation condition using a number of measurements on the order of $r$ up to some logarithmic factors. In addition, quantized measurements are further dealt in \cite{fu2018quantized} with theoretical guarantees.

\subsection{Demixing sinusoids and spikes}

Due to sensor failures or malicious environments, the measurements are susceptible to corruptions that can take arbitrary magnitudes.  
To this end, consider the problem when the observations are contaminated by sparse outliers, where
\begin{equation*}
\bm{y}_{\mathsf{corrupt}} =  \bm{x}  + \bm{s}  .
\end{equation*}
Here, $\bm{s}$ is a sparse vector, where its nonzero entries correspond to corruptions of the  observations. The goal is to decompose $\bm{x}$ and $\bm{s}$ from $\bm{y}_{\mathsf{corrupt}}$, a problem intimately related to the uncertainty principle of signal decomposition in \cite{donoho1989uncertainty,DonohoHuo01} and sparse error correction in CS \cite{li2011compressed}. 

Leveraging low-dimensional structures in both $\bx$ and $\bs$, we seek $\bm{x}$ with a small atomic norm and $\bm{s}$ with a small $\ell_1$ norm that satisfies the observation constraint \cite{fernandez2017demixing}:
\begin{equation*}
\min_{\bm{x}, \bm{s}}\; \|\bm{x}\|_{\cA} + \lambda \|\bm{s}\|_{1} \;\;\; \mbox{ subject to } \;\;\;  \bm{y}_{\mathsf{corrupt}} =  \bm{x}  + \bm{s} ,
\end{equation*}
where $\lambda>0$ is some regularization parameter. As long as the sample size is sufficiently large \cite[Theorem 2.2]{fernandez2017demixing}, and the spikes satisfy the separation condition, then the above algorithm perfectly localizes the spikes with high probability, even when the corruption amounts to near a constant fraction of the measurements.  
 
\subsection{Demixing interfering sources}

A scenario of increasing interest is when the observation is composed of a mixture of responses from multiple exciting or transmitting sources, and the goal is to simultaneously separate and localize the sources at a high resolution. For example, an electrode probing the activities in a brain records firing patterns of a few neighboring neurons, each with a distinct PSF. For pedagogical reasons, let us consider a generalization of the model \eqref{eq:samples_vector} with two interfering sources, where the observation is given as
\begin{equation*}  
\bm{y}_{\mathsf{mix}} = \mbox{diag}( \bm{g}_1 ) \bm{x}_1 + \mbox{diag}( \bm{g}_2 ) \bm{x}_2 ,
\end{equation*}
where $\bm{g}_1$ and $\bm{g}_2$ correspond to the frequency-domain response of the PSFs, and $\bm{x}_i = \sum_{k=1}^{r_i}c_{i,k} \bm{a}(\tau_{i,k})$, $i=1,2$. The goal is to separate and recover the spikes in both $\bm{x}_1$ and $\bm{x}_2$ from $\bm{y}_{\mathsf{mix}}$, where $\bm{g}_1$ and $\bm{g}_2$ are assumed known. 

Using atomic norm minimization, one seeks to recover both $\bx_1$ and $\bx_2$ simultaneously by minimizing the weighted sum of their atomic norms \cite{li2019stable}:
\begin{align*}
 \min_{\bm{x}_1, \bm{x}_2 }\quad  & \|\bm{x}_1\|_{\cA} + \lambda \|\bm{x}_2\|_{\cA} \\
  \mbox{subject to}\quad & \bm{y}_{\mathsf{mix}} = \mbox{diag}( \bm{g}_1 ) \bm{x}_1 + \mbox{diag}( \bm{g}_2 ) \bm{x}_2  ,
\end{align*}
where $\lambda>0$ is some regularization parameter. Unlike the single source case, the success of demixing critically depends on how easy it is to tell two PSFs apart -- the more similar $\bm{g}_1$ and $\bm{g}_2$ are, the harder it is to separate them. A random model can be used to generate dissimilar PSFs, namely, it is assumed the entries of $g_i$'s are i.i.d. generated from a uniform distribution over the complex circle \cite{li2019stable}. The algorithm then succeeds with high probability as long as the sample size is sufficiently large and the spikes within the same signal satisfy the separation condition \cite[Theorem 2.1]{li2019stable}, without requiring any separation for spikes coming from different sources. 

\subsection{Blind super resolution}

So far, all algorithms have assumed the PSF as known, which is a reasonable assumption for problems where one can design and calibrate the PSF a priori. In general, one might need to estimate the PSF at the same time, possibly due to the fact that the PSF may drift and needs to be calibrated on the fly during deployment. In this case, we need to revisit \eqref{eq:samples_vector} and estimate simultaneously $\bm{g}$ and $\bm{x}$ from their bilinear measurements,
$$ \bm{y}_{\mathsf{BR}} =  \mbox{diag}(\bg) \bx .$$
This problem is terribly ill-posed, as the number of unknowns far exceeds the number of observations. One remedy is to exploit additional structures of $\bm{g}$. For example, if $\bm{g}$ lies in a known low-dimensional subspace $\bm{B} =[\bm{b}_1, \cdots, \bm{b}_n]^{\top}\in\mathbb{C}^{n\times d}$ with $d\ll n$, then the degrees of freedom of $\bg$ is greatly dropped, since one only needs to estimate the coefficient $\bm{h} \in\mathbb{C}^d$ of $\bg = \bm{B}\bm{h}$ in that subspace, which is of much smaller dimension. Even such, the measurements $ \bm{y}_{\mathsf{BR}}$ is still bilinear in $\bm{h}$ and $\bm{x}$, and one cannot directly apply atomic norm minimization to $\bx$ as it does not lead to a convex program.

Interestingly, a lifting trick can be applied \cite{chi2016guaranteed}, which rewrites $\bm{y}_{\mathsf{BR}} = \mathcal{X}(\bm{Z})$ as linear measurements of a higher-dimensional object $\bm{Z} = \bm{x} \bm{h}^{\top} \in\mathbb{C}^{n\times d}$ similar to \eqref{eq:mmv_model}:
\begin{equation*}
    y_{\mathsf{BR},i} = \bm{b}_i^{\top}\bm{h} \bm{e}_i^{\top}\bm{x}  = \langle \bm{x} \bm{h}^{\top}, \bm{e}_i\bm{b}_i^{\mathsf{H}} \rangle, \quad i=1,\ldots, n,
\end{equation*}
where $\bm{e}_i$ is the $i$th standard basis vector. Consequently, one can apply atomic norm minimization to $\bm{Z}$ with respect to \eqref{eq:atomic_norm_mmv}, leading to the algorithm \cite{chi2016guaranteed}:
$$ \min_{\bm{Z}} \; \|\bm{Z}\|_{\cA}    \quad \mbox{subject to}\quad \bm{y}_{\mathsf{BR}} = \mathcal{X}(\bm{Z}). $$
This approach succeeds with high probability as soon as the sample size is sufficiently large, the spikes are well-separated and the PSF satisfies certain incoherence properties \cite[Theorem 1]{chi2016guaranteed}. Moreover, it can be further extended to demixing a mixture of sources with unknown PSFs, where each PSF lies in a distinct subspace, see \cite{yang2016super}.

\section{Beyond Line Spectrum Estimation: Super-Resolution Imaging for Single-Molecule Fluorescence Microscopy}

When the atomic set is composed of a family of complex sinusoidal signals, an {\em exact} implementation of the atomic norm in SDP is available. In the most general setting, atomic norm minimization is an infinite-dimensional convex program whose computation needs to be addressed carefully. Encouragingly, tailored solvers have been proposed and applied successfully to practical applications such as super-resolution imaging for single-molecule fluorescence microscopy, which we now present as a case study to show its promise.

\subsection{Imaging Principle}
The development of super-resolution fluorescence microscopy, which has been awarded the 2014 Nobel prize in Chemistry, is considered to fundamentally impact biological science and medicine. To date, a partial list of super-resolution fluorescence microscopy technologies includes PALM \cite{betzig2006imaging}, STORM \cite{rust2006sub}, and fPALM \cite{hess2006ultra} which share a similar imaging principle. A very nice introduction can be found at \cite{ehrenberg2014scientific}. While optical microscopy is desirable for imaging complex biological processes in live cells due to its noninvasive nature, due to diffraction limit, which is about hundreds of nanometers, it cannot image detailed internal structures of cells, which are often below 100 nanometers.

To deal with this challenge, biologists have come up with a clever idea of divide-and-conquer. To begin, imagine that every point within a cell is equipped with a photoswitchable fluorescent molecule, which means, once excited, the molecule will emit light {\em stochastically} over a duration of time to identify its location. This allows one to divide the imaging process into many frames, where in each frame, a random and sparse subset of fluorescent molecules (point sources) are activated and localized at a resolution below the diffraction limit using imaging algorithms. The final image is thus obtained by superimposing the localization outcomes of all the frames. Therefore, the high spatial resolution is achieved by sacrificing the temporal resolution. To speed up the imaging process and improve the temporal resolution, it is desirable to develop localization algorithms that are capable of identifying more fluorescent molecules per frame, which is known as the emitter density.

\begin{figure}[ht]
\begin{center}
\includegraphics[width=0.49\textwidth]{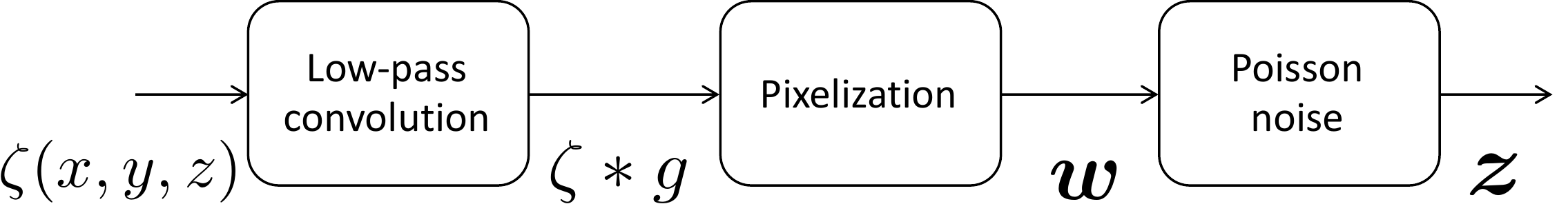}
\end{center}
\caption{The mathematical model of 3-D imaging in super-resolution fluorescence microscopy.}    \label{fig:imaging_microscopy}
\end{figure}

Very interestingly, this imaging principle can be used to reconstruct a 3-D biological structure from 2-D image frames \cite{huang2008three}. One way is to introduce a cylindrical lens to modulate the ellipticity of the PSF based on the depth of the fluorescent object, which can be modeled as a Gaussian pulse with varying ellipticity along the $x$ and $y$ directions,
$$ g (x, y | z)= \frac{1}{2 \pi \sigma_x(z) \sigma_y(z) } e^{  -\left[ \frac{x^2}{2 \sigma_x(z)^2} +\frac{y^2}{2 \sigma_y(z)^2} \right] }, $$
where $\sigma_x(z)$ and $\sigma_y(z)$ are functions of the depth in the $z$ direction, and can be calibrated in advance. For a 3-D scene of point sources,
$$\zeta (x,y,z)  = \sum_{i=1}^r c_i \delta(x-x_i, y- y_i, z-z_i), $$
its ``convolution'' with the PSF $g (x, y | z)$ is given as a 2-D image in the form of
$$ (\zeta * g)(x,y ) = \sum_{i=1}^r  \frac{c_i }{2 \pi \sigma_x(z_i) \sigma_y(z_i) } e^{  -\left[ \frac{(x-x_i) ^2}{2 \sigma_x(z_i)^2} +\frac{(y-y_i) ^2}{2 \sigma_y(z_i)^2} \right] }. $$
Therefore, to perform super-resolution, one needs to decode simultaneously the ellipticity as well as the location of the PSF, which is much more challenging. In practice, the situation is even more complex, since the continuous spatial function $(\zeta * g)(x,y ) $ needs to be further discretized due to pixelization of the detector, leading to a discretized 2-D image, $\bw  \in\mathbb{R}^{n_1\times n_2}$, where each entry of $\bw$ corresponds to the integration of $ (\zeta * g)(x,y ) $ over  the area of a pixel. The final image, $\bz$, counting the number of photons hitting the detector at each pixel, is modeled as a Poisson distribution with rate $\bw$. The whole process is summarized in Fig.~\ref{fig:imaging_microscopy}.

\subsection{Applying Atomic Norm Minimization}

Luckily, due to linearity, $\bw$ can be viewed a sparse superposition of atoms that are parameterized by the 3-D point sources,
$$ \bw = \sum_{i=1}^r c_i \ba (x_i, y_i, z_i) = \int_{x,y,z} \ba(x,y,z)d\zeta (x,y,z), $$
where each atom $\ba(x,y,z)$ corresponds to the image of a point source at $(x,y,z)$ after convolution and pixelization. The atomic set is then given as  $\cA_{\mathsf{3D}} = \{ \ba(x,y,z): x,y,z \in \mbox{imaging range} \}$.  The goal is thus to recover $\zeta(x,y,z)$, or the atomic decomposition of $\bw$, from the observation $\bz \sim \mathsf{Poisson}(\bw)$  as accurately as possible.

A natural approach is to seek the sparsest $\bw$ such that the likelihood function of the observation $\bz$ is maximized. To that end, we consider a constrained maximum likelihood estimation, where we seek to solve $\bw$ via
\begin{align}\label{eq:tvstorm}
    \min_{\bw} \quad &  - \log p (\bz | \bw )  \quad \mbox{subject to} \quad \anorm{\bw} \leq \eta,
\end{align}
where $p(\bz|\bw)$ is the Poisson likelihood function, $\|\bw\|_{\cA}$ is the induced atomic norm with respect to $\cA_{\mathsf{3D}}$, and $\eta$ is some regularization parameter that may be tuned in practice.

Early efforts such as CSSTORM \cite{zhu2012faster}, which are based on $\ell_1$ minimization by directly discretizing the parameter space, suffer from high computational complexity, due to the need of storing and manipulating a large dictionary of atoms, as fine discretization is required along all three spatial dimensions. On the other end, recent algorithmic developments such as
 ADCG \cite{boyd2017alternating} and CoGEnT \cite{rao2015forward} solve sparse inverse problems over continuous dictionaries with general convex loss functions at much reduced memory and computation requirements. In a nutshell, the ADCG method is an acceleration of conditional gradient, also known as Frank-Wolfe, to solve \eqref{eq:tvstorm} with a general atomic set $\cA = \{\ba(\theta): \theta\in \Theta \}$, where $\theta$ is a short-hand notation for the parameter space. In particular, it directly estimates the atomic decomposition of $\bw = \int \ba(\theta)d\zeta (\theta)$.

\begin{algorithm}[t]
	\caption{Alternating Descent Conditional Gradient (ADCG) \cite{boyd2017alternating}}
	\label{alg:ADCG}
	\begin{algorithmic}[H]
		\REQUIRE Observation $\bz$; Parameter $\eta>0$; \\
		\STATE Initialize $\mathcal{T}_0$ to an empty set, $\bm{c}_{0}$ to $\bm{0}$, and $j=0$; \\
		 \textbf{repeat} until stopping criteria
		\STATE Localize the next spike:
		$$\theta_{j+1} \in \argmax_{\theta \in \Theta}  \left\langle \ba(\theta_j) , \nabla_{\bw} \log p(\bz| \mathcal{A}(\mathcal{T}_j)\bc_j) \right \rangle ; $$
		\STATE Update support: $\mathcal{T}_{j+1} \gets \mathcal{T}_{j} \cup \left\{\theta_{j+1}\right\}$;
		\STATE Refinement: \textbf{repeat} \\
		 \begin{enumerate}
		 \item Update the amplitudes:
		 $$\bm{c}_{j+1} \leftarrow \argmin\limits_{\left\Vert \bm{c} \right\Vert_1 \leq \eta} \; - \log p (\bz | \mathcal{A}(\mathcal{T}_{j+1} )\bc  ) ; $$
		\item Prune support: $\mathcal{T}_{j+1} \gets \text{support}({\bm{c}}_{j+1})$;
		\item Local descent: improve $\mathcal{T}_{j+1}$ by performing local descent on $-\log p (\bz | \mathcal{A}(\mathcal{T}_{j+1} )\bc_{j+1}  ) $ holding the coefficient $\bc_{j+1}$ fixed;
		\end{enumerate}
		\STATE $j \gets j+1$;

		\ENSURE $\left(  \mathcal{T}_{j }, \bm{c}_{j} \right)$ and $\bm{x}_j = \mathcal{A}(\mathcal{T}_{j})\bm{c}_{j}$.
	\end{algorithmic}
\end{algorithm}

The standard Frank-Wolfe adds one new atom at every iteration to reduce the negative log-likelihood function, however it will introduce many spurious atoms and lose sparsity as the iteration increases. To deal with this, ADCG introduces pruning and local refinements with a hope to maintain a sparse representation at all iterations. The detailed procedure of ADCG is given in Alg.~\ref{alg:ADCG}. In the $j$th iteration, denote the support and coefficient of the current estimate of $\zeta(\theta)$ as $\mathcal{T}_j$ and $\bc_j$, and the current estimate of $\bw$ as $\mathcal{A}(\mathcal{T}_j)\bc_j$. Like Frank-Wolfe, ADCG starts by adding a spike to the estimated support $\mathcal{T}_j$ that maximally correlates with the derivative of $\log p(\bz|\bw)$ with respect to $\bw$ at the current estimate, $\mathcal{A}(\mathcal{T}_j)\bm{c}_j$. Because the spike location will be refined next, in practice, this step can be solved approximately by searching over a coarse grid of $\Theta$ to save computation.

ADCG then deviates from the standard Frank-Wolfe, and tries to improve the updated estimate by performing alternating descent over the coefficient and the support. It iterates between coefficients update via $\ell_1$ minimization, support pruning, and local refinement of the support by holding the coefficients fixed. The last step leverages the fact that $\bm{a}(\theta)$ is differentiable with respect to $\theta$, and a simple local search via gradient descent allows one to adjust the support to further reduce the loss function. Despite the nonconvexity in this refine step, \cite{boyd2017alternating} guarantees the convergence of ADCG with a convergence rate of $O(1/\epsilon)$ to reach $\epsilon$-accuracy in the function value, under some technical assumptions. In practice, the main computational benefits of ADCG are the absence of semidefinite constraints and small memory footprints, making it highly suitable for large-scale implementations.

\begin{figure}[ht]
\begin{center}
\includegraphics[width=0.49\textwidth]{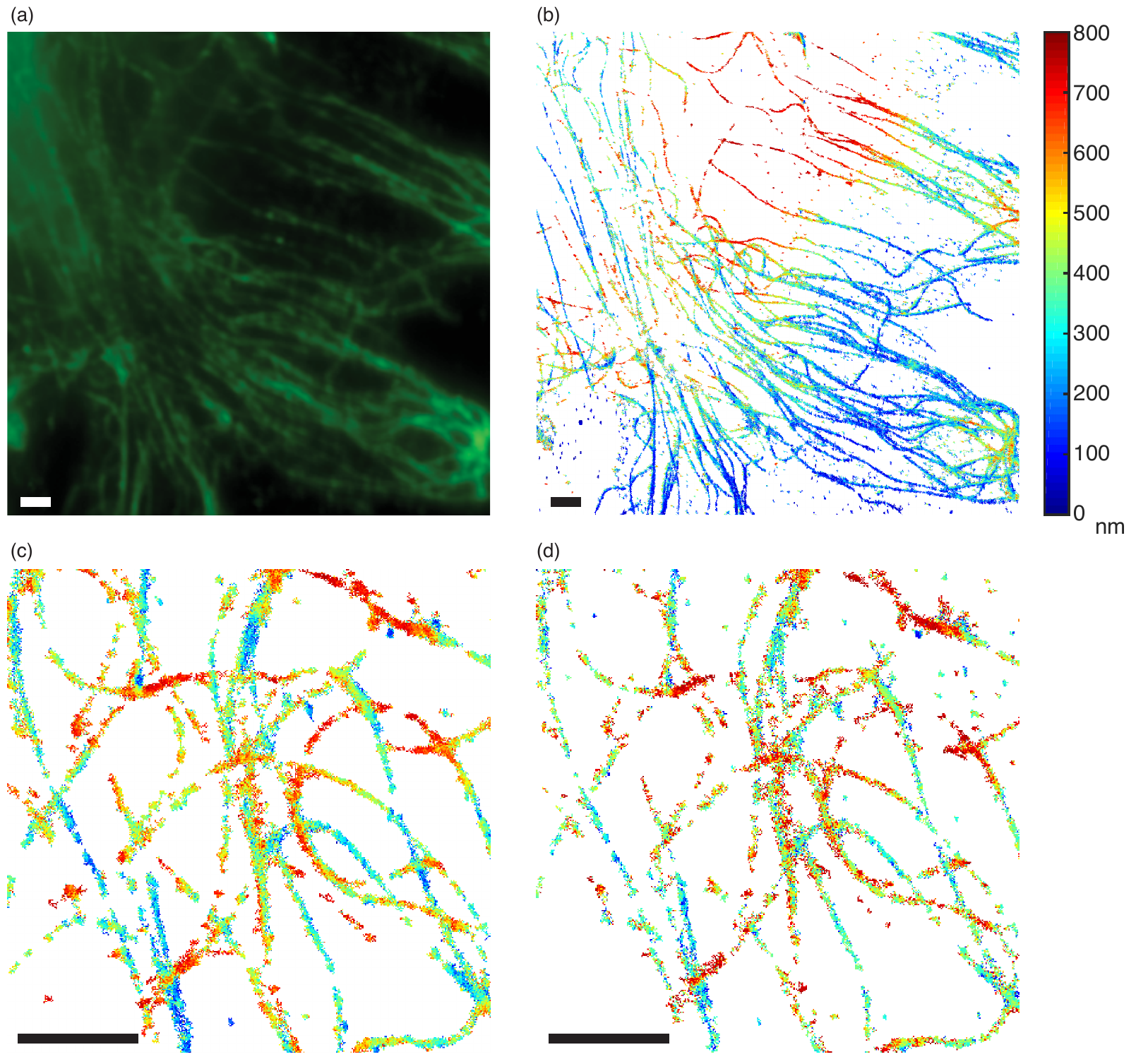}
\end{center}
\caption{(a) Diffraction-limited imaging of microtubules using conventional microscopy and (b) super-resolution 3D image reconstruction using TVSTORM. Comparison of reconstruction quality between (c) TVSTORM and (d) CSSTORM (bar: 1.5 $\mu m$). Image credit: \cite{huang2017super}.}
\label{real_3D}
\end{figure}

\begin{figure}[ht]
\begin{center}
\includegraphics[width=0.49\textwidth]{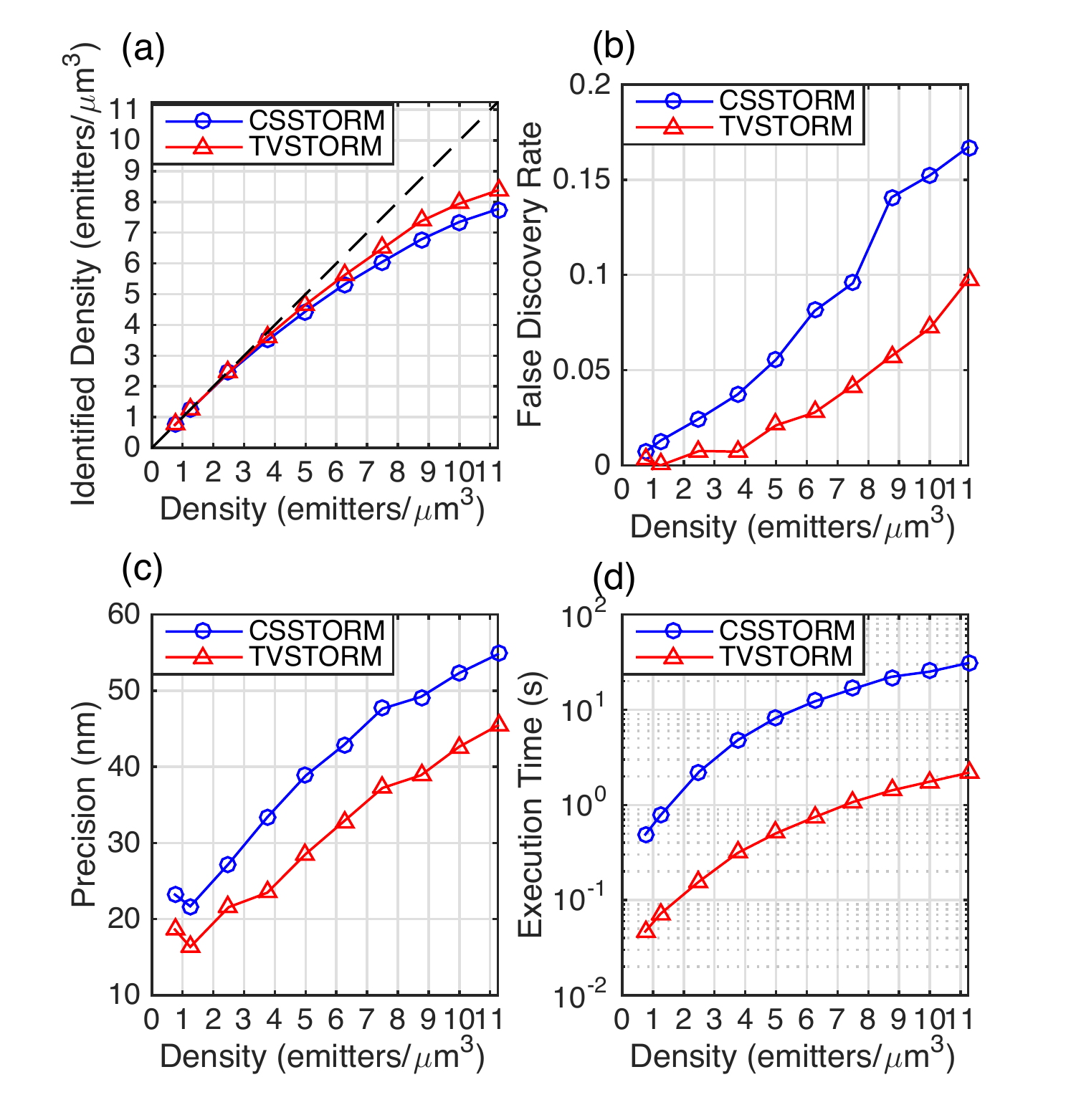}
\end{center}
\caption{Comparisons between TVSTORM
and CSSTORM for various performance metrics of 3D image reconstruction: (a) identified density, (b) false discovery rate, (c)
precision and (d) execution time with respect to the emitter density. Image credit: \cite{huang2017super}.}
\label{fig:comparison}
\end{figure}

TVSTORM \cite{huang2017super} is a modification of ADCG tailored to solve \eqref{eq:tvstorm} for 3-D image reconstruction with some domain adaptations to speed up implementations. TVSTORM outperforms CSSTORM both in terms of computational time and reconstruction quality. Fig.~\ref{real_3D} (a) shows the diffraction-limited imaging using conventional microscopy; in contrast, the 3-D super-resolution image reconstructed using TVSTORM in Fig.~\ref{real_3D} (b) is much clearer, where the structure of 3D microtubules can be well resolved with the axial coordinate represented in different colors. Fig.~\ref{real_3D} (c) and (d) compare the reconstruction quality of a zoom-in region between TVSTORM and CSSTORM, where TVSTORM provides a visually more smooth reconstruction of the line structures in microtubules. Fig.~\ref{fig:comparison} shows that TVSTORM indeed has a higher detection rate and a lower false discovery rate than CSSTORM, while executes much faster.

\section{Final Remarks}

In this paper, we presented an overview on how to leverage sparsity for continuous parameter estimation via the mathematical concept of atomic norms, which can be regarded as a generalization of the principle of $\ell_1$ norms for discrete model selection. We showcased its application in super resolution from low-pass observations in single-molecule fluorescence microscopy. The appeals of the atomic norm approach stem from its elegant mathematical framework, strong performance guarantees, and promises to scalable numerical solvers.

The atomic norm is only one of many possible approaches to exploit sparsity over the continuum. One competitive alternative is structured low-rank matrix optimization \cite{chen2013robust,jin2016general,ongie2017convex,cai2016robust}; see \cite{chi2017convex} for its connections and comparisons with the atomic norm approach. Another line of work \cite{puy2017recipes,adcock2016generalized,adcock2017breaking} generalizes the traditional CS to 
an infinite-dimensional Hilbert space. In addition, sampling theorems are developed for signals with {\em a finite rate of innovations} together with strategies for perfect reconstruction \cite{vetterli2002sampling}. More recently, a sparse functional framework has been proposed as a variational approach to handle sparsity over continuous and possibly nonlinear dictionaries. This category of estimators aims to recover functions with minimum support measure subject to the observation constraint \cite{chamon2018strong,chamon2018functional}.

As a topic still under development, open problems abound for both theoretical and practical performance of optimization-based super resolution in general. We conclude by outlining some exciting future directions.
\begin{itemize}
\item {\em Tight performance analysis in noise.} Existing analyses of atomic norm denoising \eqref{eq:DenoisingAlgorithm} typically only produce bounds that are tight up to some constant, making it less useful in practice. For example, attempts to benchmarking the theoretical bounds against the Cram\'er-Rao bound will be in vein due to the presence of large constants. It is therefore desirable to obtain tight performance bounds such as the one available for matrix denoising \cite{donoho2014minimax} that is asymptotically exact.

\item {\em Adaptive selection of regularization parameters.} One benefit of atomic norm minimization over traditional spectrum estimation approaches is that it can automatically select the model order. However, the choice of the regularization parameter   \eqref{eq:DenoisingAlgorithm} depends on the noise level, which may not be available in practice. How to optimally set the regularization parameters is another problem of great importance, see \cite{boyer2017adapting} for some recent development.

\item {\em Low-rank factorization for SDP formulations of atomic norms.} A popular heuristic to SDP with low-rank solutions is to apply low-rank matrix factorization and solve the corresponding nonconvex optimization problem \cite{burer2003nonlinear,chi2019nonconvex}, with the premise of greatly reducing its computational cost. It will be interesting to see if this approach can be applied to speed up the computation of atomic norms with performance guarantees.
  
\item {\em Bridging classical and optimization-based approaches.} There are deep connections between traditional approaches (e.g. Prony, MUSIC, and so on) and optimization-based approaches (e.g. atomic norm minimization and nuclear norm minimization). Such connections have already been realized, for instance, in the early works of Fuchs \cite{fuchs1997extension}, where he provided an optimization interpretation of the Pisarenko method \cite{pisarenko1973retrieval}. Another recent work \cite{li2018optimization} provided an optimization view to the MUSIC algorithm. It is hopeful that a confluence of past and current ideas will likely deepen our understandings and lead to further algorithmic improvements.  

\item{\em Atomic norm minimization for more general measurement models.} 
While there have been significant advances in the understanding of atomic norm minimization for line spectrum estimation, its theory and application to other measurement settings require further investigation.   
 
\item {\em Applications in communications, sensing, and imaging.} Atomic norm minimization has emerged as a popular approach recently for many practical applications, such as channel estimation in massive MIMO \cite{deng2018mmwave,tsai2018millimeter}, radar imaging \cite{zhu2016super}, and nuclear magnetic resonance (NMR) spectroscopy \cite{ying2017hankel}. It is our hope that this paper will stir more interest in applying the atomic norm in applications that call for high-resolution parameter estimation.

 \end{itemize}
 
 \section*{Acknowledgement}
 
 The authors thank the editorial board for early feedbacks of the white paper, and the associate editor and anonymous reviewers for constructive suggestions that help improve the quality of this paper. This work is supported in part by ONR under grants N00014-18-1-2142 and N00014-19-1-2404, and by NSF under grants CIF-1826519 and ECCS-1818571.

\bibliographystyle{IEEEtran}
\bibliography{Atomic.bib}

\begin{thebibliography}{100}
\providecommand{\url}[1]{#1}
\csname url@samestyle\endcsname
\providecommand{\newblock}{\relax}
\providecommand{\bibinfo}[2]{#2}
\providecommand{\BIBentrySTDinterwordspacing}{\spaceskip=0pt\relax}
\providecommand{\BIBentryALTinterwordstretchfactor}{4}
\providecommand{\BIBentryALTinterwordspacing}{\spaceskip=\fontdimen2\font plus
\BIBentryALTinterwordstretchfactor\fontdimen3\font minus
  \fontdimen4\font\relax}
\providecommand{\BIBforeignlanguage}[2]{{%
\expandafter\ifx\csname l@#1\endcsname\relax
\typeout{** WARNING: IEEEtran.bst: No hyphenation pattern has been}%
\typeout{** loaded for the language `#1'. Using the pattern for}%
\typeout{** the default language instead.}%
\else
\language=\csname l@#1\endcsname
\fi
#2}}
\providecommand{\BIBdecl}{\relax}
\BIBdecl

\bibitem{stoica1997introduction}
P.~Stoica and R.~L. Moses, \emph{Introduction to spectral analysis}.\hskip 1em
  plus 0.5em minus 0.4em\relax New Jersey: Prentice Hall, 1997, vol.~1.

\bibitem{kay1981spectrum}
S.~M. Kay and S.~L. Marple, ``Spectrum analysis—a modern perspective,''
  \emph{Proceedings of the IEEE}, vol.~69, no.~11, pp. 1380--1419, 1981.

\bibitem{krim1996two}
H.~Krim and M.~Viberg, ``Two decades of array signal processing research: the
  parametric approach,'' \emph{IEEE Signal Processing Magazine}, vol.~13,
  no.~4, pp. 67--94, 1996.

\bibitem{prony1795essai}
R.~Prony, ``Essai experimental et analytique,'' \emph{J. de l'Ecole
  Polytechnique (Paris)}, vol.~1, no.~2, pp. 24--76, 1795.

\bibitem{kumaresan1984prony}
R.~Kumaresan, D.~Tufts, and L.~L. Scharf, ``A {P}rony method for noisy data:
  Choosing the signal components and selecting the order in exponential signal
  models,'' \emph{Proceedings of the IEEE}, vol.~72, no.~2, pp. 230--233, 1984.

\bibitem{Schmidt1986MUSIC}
R.~Schmidt, ``Multiple emitter location and signal parameter estimation,''
  \emph{IEEE Transactions on Antennas and Propagation}, vol.~34, no.~3, pp.
  276--280, 1986.

\bibitem{RoyKailathESPIRIT1989}
R.~Roy and T.~Kailath, ``{ESPRIT}-estimation of signal parameters via
  rotational invariance techniques,'' \emph{IEEE Transactions on Acoustics,
  Speech and Signal Processing}, vol.~37, no.~7, pp. 984 --995, Jul 1989.

\bibitem{HuaSarkar1990}
Y.~Hua and T.~K. Sarkar, ``Matrix pencil method for estimating parameters of
  exponentially damped/undamped sinusoids in noise,'' \emph{IEEE Transactions
  on Acoustics, Speech and Signal Processing}, vol.~38, no.~5, pp. 814 --824,
  may 1990.

\bibitem{stoica1989maximum}
P.~Stoica, R.~L. Moses, B.~Friedlander, and T.~Soderstrom, ``Maximum likelihood
  estimation of the parameters of multiple sinusoids from noisy measurements,''
  \emph{IEEE Transactions on Acoustics, Speech, and Signal Processing},
  vol.~37, no.~3, pp. 378--392, 1989.

\bibitem{clark1994two}
M.~P. Clark and L.~L. Scharf, ``Two-dimensional modal analysis based on maximum
  likelihood,'' \emph{IEEE Transactions on Signal Processing}, vol.~42, no.~6,
  pp. 1443--1452, 1994.

\bibitem{CandesFernandez2012SR}
E.~J. Cand{\`e}s and C.~Fernandez-Granda, ``Towards a mathematical theory of
  super-resolution,'' \emph{Communications on Pure and Applied Mathematics},
  vol.~67, no.~6, pp. 906--956, 2014.

\bibitem{chandrasekaran2012convex}
V.~Chandrasekaran, B.~Recht, P.~A. Parrilo, and A.~S. Willsky, ``The convex
  geometry of linear inverse problems,'' \emph{Foundations of Computational
  Mathematics}, vol.~12, no.~6, pp. 805--849, 2012.

\bibitem{CheDonSau01}
S.~Chen, D.~L. Donoho, and M.~A. Saunders, ``Atomic decomposition by basis
  pursuit,'' \emph{SIAM Review}, vol.~43, no.~1, pp. 129--159, 2001.

\bibitem{recht2010guaranteed}
B.~Recht, M.~Fazel, and P.~A. Parrilo, ``Guaranteed minimum-rank solutions of
  linear matrix equations via nuclear norm minimization,'' \emph{SIAM Review},
  vol.~52, no.~3, pp. 471--501, 2010.

\bibitem{tang2012compressed}
G.~Tang, B.~N. Bhaskar, P.~Shah, and B.~Recht, ``Compressed sensing off the
  grid,'' \emph{IEEE Transactions on Information Theory}, vol.~59, no.~11, pp.
  7465--7490, 2013.

\bibitem{huang2009super}
B.~Huang, M.~Bates, and X.~Zhuang, ``Super-resolution fluorescence
  microscopy,'' \emph{Annual Review of Biochemistry}, vol.~78, pp. 993--1016,
  2009.

\bibitem{DonohoHuo01}
D.~Donoho and X.~Huo, ``Uncertainty principles and ideal atomic
  decomposition,'' \emph{IEEE Transactions on Information Theory}, vol.~47,
  no.~7, pp. 2845 --2862, 2001.

\bibitem{chen2018harnessing}
Y.~Chen and Y.~Chi, ``Harnessing structures in big data via guaranteed low-rank
  matrix estimation: Recent theory and fast algorithms via convex and nonconvex
  optimization,'' \emph{IEEE Signal Processing Magazine}, vol.~35, no.~4, pp.
  14--31, 2018.

\bibitem{CandesRombergTao2006Stable}
E.~J. Cand\`es, J.~K. Romberg, and T.~Tao, ``Stable signal recovery from
  incomplete and inaccurate measurements,'' \emph{Communications on Pure and
  Applied Mathematics}, vol.~59, no.~8, pp. 1207--1223, 2006.

\bibitem{Donoho06}
D.~Donoho, ``For most large underdetermined systems of linear equations the
  minimal $\ell_1$-norm solution is also the sparsest solution,''
  \emph{Communications on Pure and Applied Mathematics}, vol.~59, no.~6, pp.
  797--829, 2006.

\bibitem{caratheodory1911variabilitatsbereich}
C.~Carath{\'e}odory, ``{\"U}ber den variabilit{\"a}tsbereich der koeffizienten
  von potenzreihen, die gegebene werte nicht annehmen,'' \emph{Mathematische
  Annalen}, vol.~64, no.~1, pp. 95--115, 1907.

\bibitem{foucart2013mathematical}
S.~Foucart and H.~Rauhut, \emph{A Mathematical Introduction to Compressive
  Sensing}.\hskip 1em plus 0.5em minus 0.4em\relax Springer Science \& Business
  Media, 2013.

\bibitem{sanyal2011orbitopes}
R.~Sanyal, F.~Sottile, and B.~Sturmfels, ``Orbitopes,'' \emph{Mathematika},
  vol.~57, no.~2, pp. 275--314, 2011.

\bibitem{lasserre2010moments}
J.-B. Lasserre, \emph{Moments, positive polynomials and their
  applications}.\hskip 1em plus 0.5em minus 0.4em\relax World Scientific, 2010,
  vol.~1.

\bibitem{dumitrescu2007positive}
B.~Dumitrescu, \emph{Positive trigonometric polynomials and signal processing
  applications}.\hskip 1em plus 0.5em minus 0.4em\relax Springer, 2007, vol.
  103.

\bibitem{georgiou2007caratheodory}
T.~T. Georgiou, ``The {Carath{\'e}odory--Fej{\'e}r--Pisarenko} decomposition
  and its multivariable counterpart,'' \emph{IEEE Transactions on Automatic
  Control}, vol.~52, no.~2, pp. 212--228, 2007.

\bibitem{grant2008cvx}
M.~Grant and S.~Boyd, ``{CVX}: Matlab software for disciplined convex
  programming,'' available at \url{http://cvxr.com/cvx}.

\bibitem{teke2017role}
O.~Teke and P.~P. Vaidyanathan, ``On the role of the bounded lemma in the {SDP}
  formulation of atomic norm problems,'' \emph{IEEE Signal Processing Letters},
  vol.~24, no.~7, pp. 972--976, 2017.

\bibitem{choi1995sums}
M.-D. Choi, T.~Y. Lam, and B.~Reznick, ``Sums of squares of real polynomials,''
  in \emph{Proceedings of Symposia in Pure mathematics}, vol.~58, 1995, pp.
  103--126.

\bibitem{candes2008restricted}
E.~J. Cand{\`e}s, ``The restricted isometry property and its implications for
  compressed sensing,'' \emph{Comptes Rendus Mathematique}, vol. 346, no.~9,
  pp. 589--592, 2008.

\bibitem{TroppGilbert}
J.~A. Tropp and A.~C. Gilbert, ``Signal recovery from random measurements via
  orthogonal matching pursuit,'' \emph{IEEE Transactions on Information
  Theory}, vol.~53, no.~12, pp. 4655--4666, 2007.

\bibitem{Fernandez-Granda2016sr}
C.~Fernandez-Granda, ``Super-resolution of point sources via convex
  programming,'' \emph{Information and Inference: A Journal of the IMA},
  vol.~5, no.~3, pp. 251--303, 2016.

\bibitem{Ferreira2018}
M.~F. Da~Costa and W.~Dai, ``A tight converse to the spectral resolution limit
  via convex programming,'' in \emph{2018 IEEE International Symposium on
  Information Theory (ISIT)}.\hskip 1em plus 0.5em minus 0.4em\relax IEEE,
  2018, pp. 901--905.

\bibitem{moitra2015super}
A.~Moitra, ``Super-resolution, extremal functions and the condition number of
  {V}andermonde matrices,'' in \emph{Proceedings of the forty-seventh annual
  ACM symposium on Theory of computing}, 2015, pp. 821--830.

\bibitem{bhaskar2013atomic}
B.~N. Bhaskar, G.~Tang, and B.~Recht, ``Atomic norm denoising with applications
  to line spectral estimation,'' \emph{IEEE Transactions on Signal Processing},
  vol.~61, no.~23, pp. 5987--5999, 2013.

\bibitem{Tib96}
R.~Tibshirani, ``Regression shrinkage and selection via the lasso,''
  \emph{Journal of the Royal Statistical Society. Series B (Methodological)},
  vol.~58, no.~1, pp. 267--288, 1996.

\bibitem{tang2015near}
G.~Tang, B.~N. Bhaskar, and B.~Recht, ``Near minimax line spectral
  estimation,'' \emph{IEEE Transactions on Information Theory}, vol.~61, no.~1,
  pp. 499--512, 2015.

\bibitem{fernandez2013support}
C.~Fernandez-Granda, ``Support detection in super-resolution,'' in
  \emph{Proceedings of the 10th International Conference on Sampling Theory and
  Applications (SampTA 2013)}, 2013, pp. 145--148.

\bibitem{li2018approximate}
Q.~Li and G.~Tang, ``Approximate support recovery of atomic line spectral
  estimation: A tale of resolution and precision,'' \emph{Applied and
  Computational Harmonic Analysis}, 2018, in press.

\bibitem{Duval2015exact}
V.~Duval and G.~Peyr{\'e}, ``Exact support recovery for sparse spikes
  deconvolution,'' \emph{Foundations of Computational Mathematics}, vol.~15,
  no.~5, pp. 1315--1355, 2015.

\bibitem{da2019stable}
M.~F. Da~Costa and Y.~Chi, ``On the stable resolution limit of total variation
  regularization for spike deconvolution,'' \emph{arXiv preprint
  arXiv:1910.01629}, 2019.

\bibitem{Blu2008Sparse}
T.~{Blu}, P.~{Dragotti}, M.~{Vetterli}, P.~{Marziliano}, and L.~{Coulot},
  ``Sparse sampling of signal innovations,'' \emph{IEEE Signal Processing
  Magazine}, vol.~25, no.~2, pp. 31--40, March 2008.

\bibitem{rao1989performance}
B.~D. Rao and K.~S. Hari, ``Performance analysis of root-{MUSIC},'' \emph{IEEE
  Transactions on Acoustics, Speech, and Signal Processing}, vol.~37, no.~12,
  pp. 1939--1949, 1989.

\bibitem{Boyd2010ADMM}
S.~Boyd, N.~Parikh, E.~Chu, B.~Peleato, and J.~Eckstein, ``Distributed
  optimization and statistical learning via the alternating direction method of
  multipliers,'' \emph{Foundations and Trends in Machine Learning}, vol.~3,
  no.~1, pp. 1--122, 2010.

\bibitem{Golub2000Eigenvalue}
G.~H. Golub and H.~A. van~der Vorst, ``Eigenvalue computation in the 20th
  century,'' \emph{Journal of Computational and Applied Mathematics}, vol. 123,
  no.~1, pp. 35 -- 65, 2000.

\bibitem{CandesTao2006}
E.~J. Cand\`es and T.~Tao, ``Near-optimal signal recovery from random
  projections: Universal encoding strategies?'' \emph{IEEE Transactions on
  Information Theory}, vol.~52, no.~12, pp. 5406--5425, 2006.

\bibitem{Donoho2006}
D.~L. Donoho, ``Compressed sensing,'' \emph{IEEE Transactions on Information
  Theory}, vol.~52, no.~4, pp. 1289 -- 1306, 2006.

\bibitem{Tang2013sparserecovery}
G.~Tang, B.~N. Bhaskar, and B.~Recht, ``Sparse recovery over continuous
  dictionaries-just discretize,'' in \emph{2013 Asilomar Conference on Signals,
  Systems and Computers}.\hskip 1em plus 0.5em minus 0.4em\relax IEEE, 2013,
  pp. 1043--1047.

\bibitem{duval2017sparse_partI}
V.~Duval and G.~Peyr{\'e}, ``Sparse regularization on thin grids {I}: the
  {LASSO},'' \emph{Inverse Problems}, vol.~33, no.~5, p. 055008, 2017.

\bibitem{duval2017sparse_partII}
------, ``Sparse spikes super-resolution on thin grids {II}: the continuous
  basis pursuit,'' \emph{Inverse Problems}, vol.~33, no.~9, p. 095008, 2017.

\bibitem{Chi2011sensitivity}
Y.~Chi, L.~Scharf, A.~Pezeshki, and A.~Calderbank, ``Sensitivity to basis
  mismatch in compressed sensing,'' \emph{IEEE Transactions on Signal
  Processing}, vol.~59, no.~5, pp. 2182--2195, May 2011.

\bibitem{tan2014joint}
Z.~Tan, P.~Yang, and A.~Nehorai, ``Joint sparse recovery method for compressed
  sensing with structured dictionary mismatches,'' \emph{IEEE Transactions on
  Signal Processing}, vol.~62, no.~19, pp. 4997--5008, 2014.

\bibitem{mamandipoor2016newtonized}
B.~Mamandipoor, D.~Ramasamy, and U.~Madhow, ``Newtonized orthogonal matching
  pursuit: Frequency estimation over the continuum,'' \emph{IEEE Transactions
  on Signal Processing}, vol.~64, no.~19, pp. 5066--5081, 2016.

\bibitem{donoho2005sparse}
D.~L. Donoho and J.~Tanner, ``Sparse nonnegative solution of underdetermined
  linear equations by linear programming,'' \emph{Proceedings of the National
  Academy of Sciences of the United States of America}, vol. 102, no.~27, pp.
  9446--9451, 2005.

\bibitem{fuchs2005sparsity}
J.-J. Fuchs, ``Sparsity and uniqueness for some specific under-determined
  linear systems,'' in \emph{Proceedings of IEEE International Conference on
  Acoustics, Speech, and Signal Processing}, vol.~5.\hskip 1em plus 0.5em minus
  0.4em\relax IEEE, 2005, pp. v/729--v/732.

\bibitem{denoyelle2017support}
Q.~Denoyelle, V.~Duval, and G.~Peyr{\'e}, ``Support recovery for sparse
  super-resolution of positive measures,'' \emph{Journal of Fourier Analysis
  and Applications}, vol.~23, no.~5, pp. 1153--1194, 2017.

\bibitem{schiebinger2017superresolution}
G.~Schiebinger, E.~Robeva, and B.~Recht, ``Superresolution without
  separation,'' \emph{Information and Inference: A Journal of the IMA}, vol.~7,
  no.~1, pp. 1--30, 2017.

\bibitem{eftekhari2018sparse}
A.~Eftekhari, J.~Tanner, A.~Thompson, B.~Toader, and H.~Tyagi, ``Sparse
  non-negative super-resolution--simplified and stabilised,'' \emph{Applied and
  Computational Harmonic Analysis}, 2019, in press.

\bibitem{morgenshtern2016super}
V.~I. Morgenshtern and E.~J. Candes, ``Super-resolution of positive sources:
  The discrete setup,'' \emph{SIAM Journal on Imaging Sciences}, vol.~9, no.~1,
  pp. 412--444, 2016.

\bibitem{chi2015compressive}
Y.~Chi and Y.~Chen, ``Compressive two-dimensional harmonic retrieval via atomic
  norm minimization,'' \emph{IEEE Transactions on Signal Processing}, vol.~63,
  no.~4, pp. 1030--1042, 2015.

\bibitem{xu2014precise}
W.~Xu, J.-F. Cai, K.~V. Mishra, M.~Cho, and A.~Kruger, ``Precise semidefinite
  programming formulation of atomic norm minimization for recovering
  d-dimensional ($d\geq 2$) off-the-grid frequencies,'' in \emph{Information
  Theory and Applications Workshop (ITA), 2014}.\hskip 1em plus 0.5em minus
  0.4em\relax IEEE, 2014, pp. 1--4.

\bibitem{li2016offgrid}
Y.~Li and Y.~Chi, ``Off-the-grid line spectrum denoising and estimation with
  multiple measurement vectors,'' \emph{IEEE Transactions on Signal
  Processing}, vol.~64, no.~5, pp. 1257--1269.

\bibitem{tian2007compressed}
Z.~Tian and G.~Giannakis, ``Compressed sensing for wideband cognitive radios,''
  in \emph{Acoustics, Speech and Signal Processing, 2007. ICASSP 2007. IEEE
  International Conference on}, vol.~4.\hskip 1em plus 0.5em minus 0.4em\relax
  IEEE, 2007, pp. 1357--1360.

\bibitem{candes2011probabilistic}
E.~J. Cand\`es and Y.~Plan, ``A probabilistic and {RIP}less theory of
  compressed sensing,'' \emph{IEEE Transactions on Information Theory},
  vol.~57, no.~11, pp. 7235--7254, 2011.

\bibitem{heckel2018generalized}
R.~Heckel and M.~Soltanolkotabi, ``Generalized line spectral estimation via
  convex optimization,'' \emph{IEEE Transactions on Information Theory},
  vol.~64, no.~6, pp. 4001--4023, 2018.

\bibitem{fu2018quantized}
H.~Fu and Y.~Chi, ``Quantized spectral compressed sensing: Cramer--rao bounds
  and recovery algorithms,'' \emph{IEEE Transactions on Signal Processing},
  vol.~66, no.~12, pp. 3268--3279, 2018.

\bibitem{donoho1989uncertainty}
D.~L. Donoho and P.~B. Stark, ``Uncertainty principles and signal recovery,''
  \emph{SIAM Journal on Applied Mathematics}, vol.~49, no.~3, pp. 906--931,
  1989.

\bibitem{li2011compressed}
X.~Li, ``\BIBforeignlanguage{English}{Compressed sensing and matrix completion
  with constant proportion of corruptions},''
  \emph{\BIBforeignlanguage{English}{Constructive Approximation}}, vol.~37, pp.
  73--99, 2013.

\bibitem{fernandez2017demixing}
C.~Fernandez-Granda, G.~Tang, X.~Wang, and L.~Zheng, ``Demixing sines and
  spikes: Robust spectral super-resolution in the presence of outliers,''
  \emph{Information and Inference: A Journal of the IMA}, vol.~7, no.~1, pp.
  105--168, 2017.

\bibitem{li2019stable}
Y.~Li and Y.~Chi, ``Stable separation and super-resolution of mixture models,''
  \emph{Applied and Computational Harmonic Analysis}, vol.~46, no.~1, pp.
  1--39, 2019.

\bibitem{chi2016guaranteed}
Y.~Chi, ``Guaranteed blind sparse spikes deconvolution via lifting and convex
  optimization,'' \emph{IEEE Journal of Selected Topics in Signal Processing},
  vol.~10, no.~4, pp. 782--794, 2016.

\bibitem{yang2016super}
D.~Yang, G.~Tang, and M.~B. Wakin, ``Super-resolution of complex exponentials
  from modulations with unknown waveforms,'' \emph{IEEE Transactions on
  Information Theory}, vol.~62, no.~10, pp. 5809--5830, 2016.

\bibitem{betzig2006imaging}
E.~Betzig, G.~H. Patterson, R.~Sougrat, O.~W. Lindwasser, S.~Olenych, J.~S.
  Bonifacino, M.~W. Davidson, J.~Lippincott-Schwartz, and H.~F. Hess, ``Imaging
  intracellular fluorescent proteins at nanometer resolution,'' \emph{Science},
  vol. 313, no. 5793, pp. 1642--1645, 2006.

\bibitem{rust2006sub}
M.~J. Rust, M.~Bates, and X.~Zhuang, ``Sub-diffraction-limit imaging by
  stochastic optical reconstruction microscopy ({STORM}),'' \emph{Nature
  Methods}, vol.~3, no.~10, pp. 793--796, 2006.

\bibitem{hess2006ultra}
S.~T. Hess, T.~P. Girirajan, and M.~D. Mason, ``Ultra-high resolution imaging
  by fluorescence photoactivation localization microscopy,'' \emph{Biophysical
  journal}, vol.~91, no.~11, pp. 4258--4272, 2006.

\bibitem{ehrenberg2014scientific}
\BIBentryALTinterwordspacing
M.~Ehrenberg, ``Scientific background on the {Nobel Prize in Chemistry} 2014,''
  2014. [Online]. Available:
  \url{www.nobelprize.org/nobel_prizes/chemistry/laureates/2014/advanced-chemistryprize2014.pdf}
\BIBentrySTDinterwordspacing

\bibitem{huang2008three}
B.~Huang, W.~Wang, M.~Bates, and X.~Zhuang, ``Three-dimensional
  super-resolution imaging by stochastic optical reconstruction microscopy,''
  \emph{Science}, vol. 319, no. 5864, pp. 810--813, 2008.

\bibitem{zhu2012faster}
L.~Zhu, W.~Zhang, D.~Elnatan, and B.~Huang, ``Faster {STORM} using compressed
  sensing,'' \emph{Nature Methods}, vol.~9, no.~7, p. 721, 2012.

\bibitem{boyd2017alternating}
N.~Boyd, G.~Schiebinger, and B.~Recht, ``The alternating descent conditional
  gradient method for sparse inverse problems,'' \emph{SIAM Journal on
  Optimization}, vol.~27, no.~2, pp. 616--639, 2017.

\bibitem{rao2015forward}
N.~Rao, P.~Shah, and S.~Wright, ``Forward--backward greedy algorithms for
  atomic norm regularization,'' \emph{IEEE Transactions on Signal Processing},
  vol.~63, no.~21, pp. 5798--5811, 2015.

\bibitem{huang2017super}
J.~Huang, M.~Sun, J.~Ma, and Y.~Chi, ``Super-resolution image reconstruction
  for high-density three-dimensional single-molecule microscopy,'' \emph{IEEE
  Transactions on Computational Imaging}, vol.~3, no.~4, pp. 763--773, 2017.

\bibitem{chen2013robust}
Y.~Chen and Y.~Chi, ``Robust spectral compressed sensing via structured matrix
  completion,'' \emph{IEEE Transactions on Information Theory}, vol.~60,
  no.~10, pp. 6576--6601, Oct 2014.

\bibitem{jin2016general}
K.~H. Jin, D.~Lee, and J.~C. Ye, ``A general framework for compressed sensing
  and parallel {MRI} using annihilating filter based low-rank {H}ankel
  matrix,'' \emph{IEEE Transactions on Computational Imaging}, vol.~2, no.~4,
  pp. 480--495, 2016.

\bibitem{ongie2017convex}
G.~Ongie, S.~Biswas, and M.~Jacob, ``Convex recovery of continuous domain
  piecewise constant images from nonuniform {F}ourier samples,'' \emph{IEEE
  Transactions on Signal Processing}, vol.~66, no.~1, pp. 236--250.

\bibitem{cai2016robust}
J.-F. Cai, X.~Qu, W.~Xu, and G.-B. Ye, ``Robust recovery of complex exponential
  signals from random gaussian projections via low rank {H}ankel matrix
  reconstruction,'' \emph{Applied and computational harmonic analysis},
  vol.~41, no.~2, pp. 470--490, 2016.

\bibitem{chi2017convex}
Y.~Chi, ``Convex relaxations of spectral sparsity for robust super-resolution
  and line spectrum estimation,'' in \emph{Wavelets and Sparsity XVII}, vol.
  10394.\hskip 1em plus 0.5em minus 0.4em\relax International Society for
  Optics and Photonics, 2017, p. 103941G.

\bibitem{puy2017recipes}
G.~Puy, M.~E. Davies, and R.~Gribonval, ``Recipes for stable linear embeddings
  from {H}ilbert spaces to $\mathbb{R}^m$,'' \emph{IEEE Transactions on
  Information Theory}, vol.~63, no.~4, pp. 2171--2187, 2017.

\bibitem{adcock2016generalized}
B.~Adcock and A.~C. Hansen, ``Generalized sampling and infinite-dimensional
  compressed sensing,'' \emph{Foundations of Computational Mathematics},
  vol.~16, no.~5, pp. 1263--1323, 2016.

\bibitem{adcock2017breaking}
B.~Adcock, A.~C. Hansen, C.~Poon, and B.~Roman, ``Breaking the coherence
  barrier: A new theory for compressed sensing,'' in \emph{Forum of
  Mathematics, Sigma}, vol.~5.\hskip 1em plus 0.5em minus 0.4em\relax Cambridge
  University Press, 2017.

\bibitem{vetterli2002sampling}
M.~Vetterli, P.~Marziliano, and T.~Blu, ``Sampling signals with finite rate of
  innovation,'' \emph{IEEE Transactions on Signal Processing}, vol.~50, no.~6,
  pp. 1417--1428, 2002.

\bibitem{chamon2018strong}
L.~F. Chamon, Y.~C. Eldar, and A.~Ribeiro, ``Strong duality of sparse
  functional optimization,'' in \emph{2018 IEEE International Conference on
  Acoustics, Speech and Signal Processing (ICASSP)}.\hskip 1em plus 0.5em minus
  0.4em\relax IEEE, 2018, pp. 4739--4743.

\bibitem{chamon2018functional}
------, ``Functional nonlinear sparse models,'' \emph{arXiv preprint
  arXiv:1811.00577}, 2018.

\bibitem{donoho2014minimax}
D.~Donoho and M.~Gavish, ``Minimax risk of matrix denoising by singular value
  thresholding,'' \emph{The Annals of Statistics}, vol.~42, no.~6, pp.
  2413--2440, 2014.

\bibitem{boyer2017adapting}
C.~Boyer, Y.~De~Castro, and J.~Salmon, ``{Adapting to unknown noise level in
  sparse deconvolution},'' \emph{Information and Inference: A Journal of the
  IMA}, vol.~6, no.~3, pp. 310--348, 2017.

\bibitem{burer2003nonlinear}
S.~Burer and R.~D. Monteiro, ``A nonlinear programming algorithm for solving
  semidefinite programs via low-rank factorization,'' \emph{Mathematical
  Programming}, vol.~95, no.~2, pp. 329--357, 2003.

\bibitem{chi2019nonconvex}
Y.~Chi, Y.~M. Lu, and Y.~Chen, ``Nonconvex optimization meets low-rank matrix
  factorization: An overview,'' \emph{IEEE Transactions on Signal Processing},
  vol.~67, no.~20, pp. 5239--5269, 2019.

\bibitem{fuchs1997extension}
J.-J. Fuchs, ``Extension of the {Pisarenko} method to sparse linear arrays,''
  \emph{IEEE Transactions on Signal Processing}, vol.~45, no.~10, pp.
  2413--2421, 1997.

\bibitem{pisarenko1973retrieval}
V.~F. Pisarenko, ``The retrieval of harmonics from a covariance function,''
  \emph{Geophysical Journal International}, vol.~33, no.~3, pp. 347--366, 1973.

\bibitem{li2018optimization}
S.~Li, H.~Mansour, and M.~B. Wakin, ``An optimization view of {MUSIC} and its
  extension to missing data,'' \emph{arXiv preprint arXiv:1806.03511}, 2018.

\bibitem{deng2018mmwave}
J.~Deng, O.~Tirkkonen, and C.~Studer, ``Mmwave channel estimation via atomic
  norm minimization for multi-user hybrid precoding,'' in \emph{2018 IEEE
  Wireless Communications and Networking Conference (WCNC)}.\hskip 1em plus
  0.5em minus 0.4em\relax IEEE, 2018, pp. 1--6.

\bibitem{tsai2018millimeter}
Y.~Tsai, L.~Zheng, and X.~Wang, ``Millimeter-wave beamformed full-dimensional
  {MIMO} channel estimation based on atomic norm minimization,'' \emph{IEEE
  Transactions on Communications}, vol.~66, no.~12, pp. 6150--6163, 2018.

\bibitem{zhu2016super}
Z.~Zhu, G.~Tang, P.~Setlur, S.~Gogineni, M.~B. Wakin, and M.~Rangaswamy,
  ``Super-resolution in {SAR} imaging: Analysis with the atomic norm,'' in
  \emph{2016 IEEE Sensor Array and Multichannel Signal Processing Workshop
  (SAM)}.\hskip 1em plus 0.5em minus 0.4em\relax IEEE, 2016, pp. 1--5.

\bibitem{ying2017hankel}
J.~Ying, H.~Lu, Q.~Wei, J.-F. Cai, D.~Guo, J.~Wu, Z.~Chen, and X.~Qu, ``Hankel
  matrix nuclear norm regularized tensor completion for $n$-dimensional
  exponential signals,'' \emph{IEEE Transactions on Signal Processing},
  vol.~65, no.~14, pp. 3702--3717, 2017.

\end{thebibliography}

\begin{IEEEbiographynophoto}{Yuejie Chi}
(S'09-M'12-SM'17)
received the Ph.D. degree in Electrical Engineering from Princeton University in 2012, and the B.E. (Hon.) degree in Electrical Engineering from Tsinghua University, Beijing, China, in 2007. She was with The Ohio State University from 2012 to 2017. Since 2018, she is an Associate Professor with the department of Electrical and Computer Engineering at Carnegie Mellon University, where she holds the Robert E. Doherty Early Career Development Professorship. Her research interests include signal processing, statistical inference, machine learning, large-scale optimization, and their applications in data science, inverse problems, imaging, and sensing systems. She is a recipient of the PECASE Award, NSF CAREER Award, AFOSR and ONR Young Investigator Program Awards, Ralph E. Powe Junior Faculty Enhancement Award, and Google Faculty Research Award. She received IEEE SPS Early Career Technical Achievement Award and Young Author Best Paper Award from IEEE Signal Processing Society, and Best Paper Award at the IEEE International Conference on Acoustics, Speech, and Signal Processing (ICASSP). She has served as an Elected Member of the SPTM, SAM and MLSP Technical Committees of the IEEE Signal Processing Society. She currently serves as an Associate Editor of IEEE Trans. on Signal Processing.
\end{IEEEbiographynophoto}

\begin{IEEEbiographynophoto}{Maxime Ferreira Da Costa}
(S'14-M'18)
received the Ph.D. degree in Electrical and Electronic Engineering from Imperial College London, UK, in 2018. He has been awarded the M.Sc in Signal Processing from Imperial College London in 2012, and the Dipl{\^o}me d'Ing{\'e}nieur from Sup{\'e}lec, France, the same year. Since October 2018, he is a Research Associate with the department of Electrical and Computer Engineering at Carnegie Mellon University. His research focuses on mathematical signal processing, inverse problems and optimization. In 2018, he was shortlisted among the finalists for the Jack Keil Wolf Student Paper Award at the IEEE International Symposium on Information Theory (ISIT).

\end{IEEEbiographynophoto}


\end{document}